\documentclass[useAMS,usenatbib]{mn2e} 
\usepackage{graphicx,subfigure}
\usepackage{pslatex}

\title[Abundances and stratification in the atmospheres of four HgMn
stars]{Abundances and search for vertical stratification in the
atmospheres of four HgMn stars}

\author[M. Thiam et al.]  {M. Thiam$^{1}$\thanks{E-mail:
emt9755@umoncton.ca}, F. LeBlanc$^{1}$, V. Khalack$^{1}$,
G.A. Wade$^{2}$\\ $^{1}$D\'epartement de Physique et d'Astronomie,
Universit\'e de Moncton, Moncton, N.-B., E1A 3E9, Canada\\
$^{2}$Department of Physics, Royal Military College of Canada, PO Box
17000 stn 'FORCES', Kingston, Ontario, K7K 7B4, Canada}

\begin{document}

\date{Accepted . Received ; in original form}

\pagerange{\pageref{firstpage}--\pageref{lastpage}} \pubyear{2010}

\maketitle

\label{firstpage}

\begin{abstract}{Using high resolution, high-S/N archival UVES spectra, we have performed
a detailed spectroscopic analysis of 4 chemically peculiar HgMn stars
(HD~71066, HD~175640, HD~178065 and HD~221507). Using spectrum
synthesis, mean photospheric chemical abundances are derived for 22
ions of 16 elements. We find good agreement between our derived
abundances and those published previously by other authors. For the 5
elements that present a sufficient number of suitable lines, we have
attempted to detect vertical chemical stratification by analyzing the
dependence of derived abundance as a function of optical depth. For
most elements and most stars we find no evidence of chemical
stratification with typical $3\sigma$ upper limits of $\Delta\log
N_{\rm elem}/N_{\rm tot}\sim 0.1-0.2$ dex per unit optical
depth. However, for Mn in the atmosphere of HD 178065 we find
convincing evidence of stratification. Modeling of the line profiles
using a two-step model for the abundance of Mn yields a local
abundance varying approximately linearly by $\sim 0.7$ dex through the
optical depth range log~$\tau_{\rm 5000}=-3.6$ to --2.8.}
\end{abstract}


\section{Introduction}

Chemically peculiar (CP) stars are A- and B-type stars exhibiting
peculiar chemical abundances and slow rotation. With effective
temperatures ranging between 10~000~K and 15~000~K, the late B-type
HgMn stars belong to the third group of CP stars defined by Preston
(1974). They are characterized spectroscopically by large
overabundances of Hg (possibly up to 6~dex; Heacox 1979, Cowley et
al. 2006) and Mn (possibly up to 3 dex; Aller 1970). Overabundances
greater than 2~dex of Be, Ga, Y, Pt, Bi, Xe, Eu, Gd, W and Pb are
frequently observed in these stars, while He, N, Mg, Al, Ni and Zn
often are underabundant by more than 0.5~dex (Takada-Hidai 1991). The
elements C, O, Si, S, Ca, V, Cr and Fe usually show only modest
peculiarities, with abundances within 0.5~dex of their solar
abundances (classified as "normal type" elements by Takada-Hidai 1991).

HgMn stars rotate slowly in comparison to "normal" late-B stars, with
$v~\sin~i<100~{\rm km~s^{-1}}$.  Their average rotation velocity is
estimated to be 29~km~s$^{-1}$ (Abt et al. 1972). These stars present
a high rate of binarity: Gerbaldi et al. (1985) estimate that 51\% are
members of binary systems. Most of these binary systems have
relatively small separations, with 41\% belonging to SB1 systems,
while 59\% belong to SB2 systems. Some authors (Mathys \& Hubrig 1995;
Hubrig et al. 1999; Hubrig \& Castelli 2001) have reported the
presence of a weak magnetic fields in the atmospheres of some HgMn
stars. However, attempts to confirm these claims for other
HgMn stars have been  unsuccessful (Landstreet 1982; Shorlin et
al. 2002; Wade et al. 2006). In light of the lack of convincing
evidence, we assume for the purposes of this study that HgMn stars are
non-magnetic objects.  Because of their high effective temperatures,
the HgMn stars should not host hydrogen convective zones in their
envelopes, and their low He abundances may reduce the importance of He
convection zones as well.  Indeed, Adelman (1994) showed that HgMn
stars have small microturbulent velocities. Atmospheric velocity
fields are estimated to be less than 1~km~s$^{-1}$ for stars with
$T_{\rm eff}$ ranging between 10~200~K and 12~000~K (Landstreet 1998).

\begin{table*}
\center
\begin{minipage}{155mm}
\caption{Journal of observations and stellar parameters of studied stars.}
\label{tab:tousensemble}
\begin{tabular}{ l l l c c c c  c c l c l l  }\hline\hline 
HD     & Other name  & HJD & Program ID & Binarity & $T_{\rm{eff}}$ & log $g$ & $V_{\rm r}$  & $v~\sin~i$     & Reference\\ 
       &      &      &            &          &  (K)           & (cgs)   & (km s$^{-1}$)& (km s$^{-1}$)  &    \\\hline 
71066  & $\kappa^2$~Vel & 2453098.50  & 60.A-9022   &     & 12 010  & 3.95    &$+16.8$  & 2.0   &  Hubrig et al. (1999)\\ 
175640 & HR~7143  & 2452074.89  & 67.D-0579   & SB1 & 11 958  & 3.95    &$-34.2$  & 2.0   &   Hubrig et al. (1999)\\ 
178065 & HR~7245 & 2452074.92  & 67.D-0579   & SB1 & 12 193  & 3.54    &$-34.8$  & 1.5   &   Hubrig et al. (1999)\\ 
221507 & $\beta$~Scl & 2452435.94  & 266.D-5655  &     & 12 476  & 4.13    &$-24.1$  & 25.0  &  Dolk et al. (2003)\\\hline
\end{tabular}
\end{minipage}
\end{table*}

The slow rotation and weak microturbulence of HgMn stars suggest that
their atmospheres should be hydrodynamically stable. Atomic diffusion
(Michaud 1970), resulting from the competition between local
gravitational settling and radiative acceleration, may therefore play
an important role in determining the chemical properties of their
outer layers. A natural consequence of diffusion is the presence of
non-uniform vertical distributions of chemical abundances (chemical
stratification).  Detecting and characterizing such stratification is
the primary goal of this study.

Several investigations aimed at the detection of chemical
stratification in atmospheres of HgMn stars were undertaken in the
past. Savanov \& Hubrig (2003) reported stratification of Cr in a
sample of 10 HgMn stars.  Their method involved the derivation of the
abundance of Cr from 8 lines of Cr~{\scriptsize{II}} multiplet 30
located in the Stark-broadened wings of the H${\beta}$ line. According
to these authors, the average chromium abundance from 9 stars of their
sample, increases toward the core of H${\beta}$, and therefore
presumably toward the upper atmosphere by $0.34\pm0.12$~dex.  The
difference of $0.2$~dex in the average abundances between
Cr~{\scriptsize{I}} and Cr~{\scriptsize{II}} in the HgMn star
HD~175640, obtained by Castelli \& Hubrig (2004), is interpreted by
these authors as a confirmation of Cr stratification found by Savanov
\& Hubrig (2003). An increase of Mn toward the upper atmosphere of
HgMn stars was reported by Alecian (1982) and Sigut (2001). Evidence
of the stratification of Ga has also been published by Lanz et
al. (1993).

Meanwhile, Dubaj et al.  (2004) modeled Fe lines in the spectrum of
the HgMn star HD~143807A shortward and longward of the Balmer
jump. They found no conclusive evidence of Fe stratification. This
suggests that if stratification of these elements is present in the
atmospheres of HgMn stars, it is relatively weak (in comparison to
that frequently observed in magnetic Ap stars, for example -
e.g. Ryabchikova et al. 2005).

Additional evidence for the presence of chemical stratification in
HgMn stars has also been published.  Sigut et al. (2000) signaled the
discovery of Mn~{\scriptsize{II}}, P~{\scriptsize{II}} and
Hg~{\scriptsize{II}} emission lines in the spectrum of the HgMn star
46 Aql.  Sigut (2001) interpreted these emission lines as the result
of interlocked non-Local Thermodynamic Equilibrium (NLTE) processes in
the presence of chemical stratification.  Wahlgren \& Hubrig (2000)
also identified several emission lines of other elements, mostly
Cr~{\scriptsize{II}} and Mn~{\scriptsize{II}}, in the spectra of some
HgMn stars. They postulated that the emissions of the aforementioned
elements in the red spectral region arise from a selective excitation
process involving Ly$\alpha$ photon energies.

Very recently, Khalack et al. (2007 and 2008) have detected
stratification of several elements including Fe in Blue Horizontal
Branch (BHB) stars, evolved objects with effective temperatures
similar to those of HgMn stars.  These authors found that the iron
abundance increases toward the lower atmosphere in three BHB stars.
They also reported nitrogen and sulfur stratification in the hot BHB
star HD~135485. In this star, the abundances of those elements
increase toward the upper atmosphere. The method used to detect
vertical stratification in BHB stars in the aforementioned papers is
applied here to HgMn stars.

The aim of this paper is to search for vertical stratification of
elements in the atmospheres of HgMn stars.  Our approach is to model a
large number of absorption lines of many different elements, and to
examine the variation of the derived abundances as a function of
optical depth. The mean abundances of the various elements analyzed
will also be obtained.

\section{Observations and spectral reduction}

Spectra of the HgMn stars HD~71066 ($\kappa^2$~Vel), HD~175640 (HR~7143),
HD~178065 (HR~7245) and HD~221507 ($\beta$~Scl) were obtained from the ESO
Science
Archive\footnote{http://archive.eso.org/eso/eso\_archive\_main.html}. The
Heliocentric Julian Date and program ID of these
stars are summarized in Table~\ref{tab:tousensemble}. The spectra were
obtained with the VLT UV-Visual Echelle Spectrograph (UVES) on the UT2
unit telescope. This instrument covers a wide spectral region from
$\lambda = 3040$~\AA~to $10~000$~\AA. The slit widths for the two arms
of UVES were $0.4\arcsec$ for the blue and $0.3\arcsec$ for the
red. The resultant resolving power is approximately
$\lambda/\Delta\lambda = 80~000$ in the blue and $110~000$ in the
red. The recorded spectra exhibit gaps which are located between
$5730-5840$~\AA~and $8510-8660$~\AA.

Spectral reduction was performed using the UVES pipeline Data
Reduction Software (version 2.2.0) for the 346 ($304-388$~nm) and 437
($373-499$~nm) settings. Each order of these spectra was normalized
using the IRAF ``continuum'' procedure. The signal-to-noise ratio
after normalization ranges between S/N = 200 and 600. For the 580
($476-684$~nm) and 860 ($660-1040$~nm) settings, we used a software
provided by  V. Tsymbal (private communication) for spectral reduction
and normalization, because it works faster and has convenient
interface.

Before adopting Tsymbal's code, we applied it to reduce the 580
settings for two stars of our sample. The same procedure was also
performed using the UVES pipeline.  A comparison of the resulting
reduced spectra yields excellent agreement.
After spectral normalization, observed and synthetic line profiles
were compared to determine the radial velocities of our sample of
stars. The radial velocity obtained for each star is reported in
col.~8 of Table~\ref{tab:tousensemble}.

\section{Spectrum synthesis}

In our analysis, we have adopted the stellar atmosphere parameters
derived for the investigated stars by Hubrig et al. (1999) and Dolk et
al. (2003). 
The adopted atmospheric and spectroscopic parameters,
including the effective temperature, surface gravity and projected
rotational velocity, are reported in
Table~\ref{tab:tousensemble}. 
The atmosphere models used for spectrum synthesis are computed with
ATLAS9 (Kurucz 1993). They are calculated with a solar metallicity and
a microturbulent velocity $\xi = 2.0$ km~s$^{-1}$. For atomic data, we
used primarily the Vienna Atomic Line
Database\footnote{http://ams.astro.univie.ac.at/vald} (VALD) (Martin et al. 1988; Kurucz 1993 and 1994; Kupka et
al. 1999; Ryabchikova et al. 1999). We also extracted some phosphorus
and mercury atomic data from the National Institute of Standards and
Technology\footnote{http://physics.nist.gov/PhysRefData/ASD/lines\_form.html}
(NIST) (Martin et al. 1985; Ralchenko et al. 2007). The NIST database
was used because VALD does not provide a large number of P and Hg
atomic data. Meanwhile, titanium atomic data from Pickering et al. (2001) have been used to compare the derived abundances from different sources. The lines and associated atomic data used in our
analysis are summarized in Table~\ref{tab:donnees_atomiques} (online material).

To determine elemental abundances and to characterize the optical
depth of line formation, we used the \textsc{Zeeman2} LTE spectrum
synthesis code (Landstreet 1988; Wade et al. 2001a).  Although
developed for spectrum synthesis of magnetic stars, this code is also
able to reproduce stellar spectra in the non-magnetic case, including
the effects of non-solar and vertically-stratified abundances.

As shown by Adelman (1994) and Landstreet (1998), and implicitly
confirmed in the study of Dolk et al. (2003), HgMn stars have small
microturbulent velocities. In this study, line profiles were
synthesized assuming $\xi = 0$ km~s$^{-1}$, an assumption that will be
verified {\em a posteriori} in Sect. 4. Since we desire high precision
results, we carefully pre-selected lines to avoid blends and
misidentifications. We also only selected lines located within
spectral orders longward of the Balmer jump, in the $437$ and $580$
settings, for which S/N~$\ge~300$. For our spectral analysis, each
line profile of a particular species is fitted independently, while
assuming a vertically-uniform abundance of the element in the
atmosphere.

\begin{figure}
\vspace{455pt} \includegraphics[scale=0.31,angle=-90]{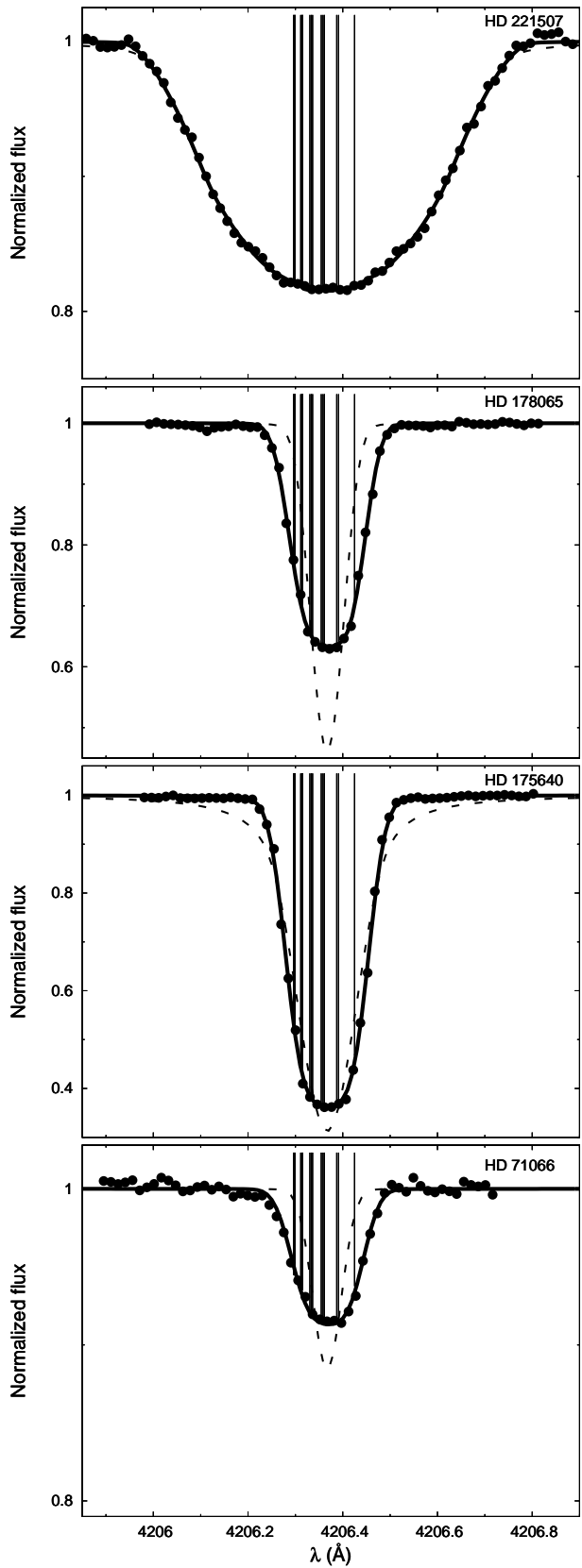}
\vspace{20pt}
\caption{The Mn~{\scriptsize{II}}~$\lambda~4206$ line observed (dots)
for HD~71066, HD~175640, HD~178065 and HD~221507. This line is fitted
by assuming a single-line component (dashed curve) and the hfs data
(solid curve). The vertical lines represent the hfs line positions.}
\label{fit:MnHFS}
\end{figure}

\begin{figure}
\vspace{455pt} \includegraphics[scale=0.31,angle=-90]{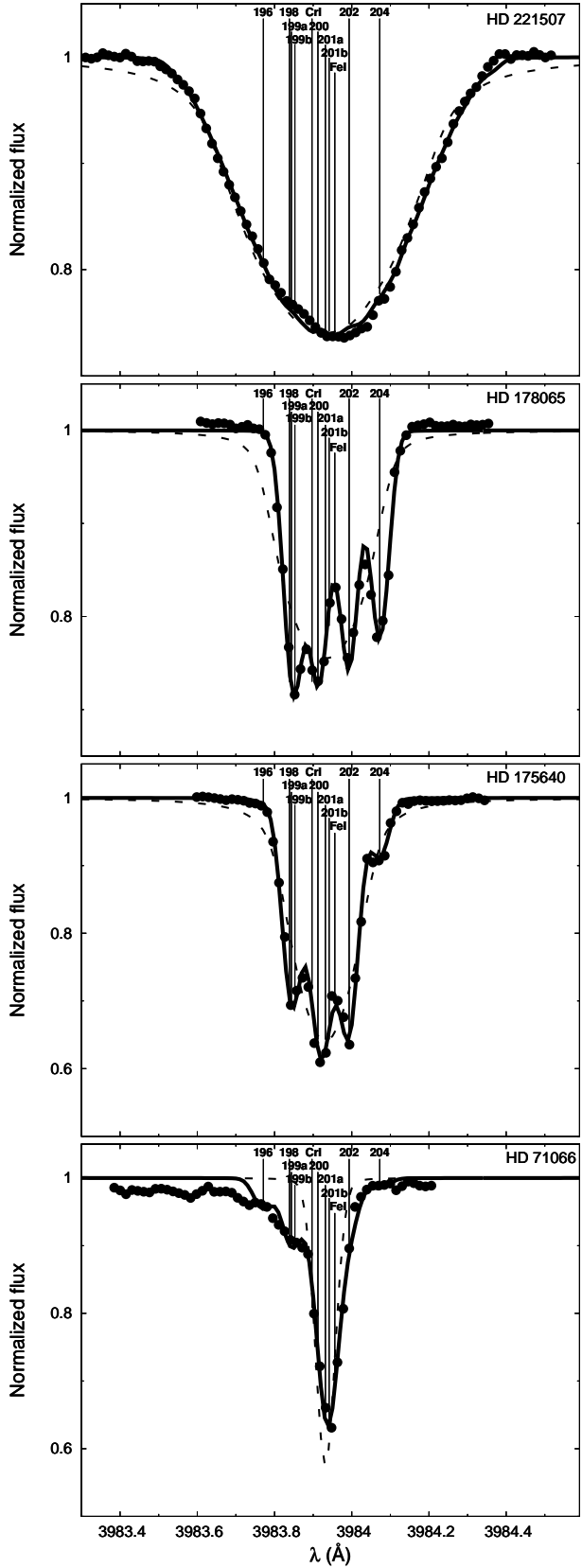}
\vspace{20pt}
\caption{Same description as in Fig.~\ref{fit:MnHFS}, but for the
observed Hg~{\scriptsize{II}}~$\lambda~3984$ line (dots), fitted by
assuming a single-line component (dashed curve) and the isotopic
composition (solid curve). The vertical lines represent the lines of
the various isotopes and blends.}
\label{fit:HgISS}
\end{figure}

Hyperfine and isotopic splitting are respectively important for Mn
(Jomaron et al. 1999) and Hg (White et al. 1976; Smith 1997; Dolk et
al. 2003) line profiles synthesis in HgMn stars. When modeling these
lines, we include the data of Mn hyperfine structure (hfs) and Hg
isotopic splitting, when they are available. The data for Mn hyperfine
structure are taken from Holt et al. (1999), while those of the Hg
isotopic splitting are from Dolk et al. (2003). These data are
summarized respectively in Table~\ref{tab:MnHFS} and
Table~\ref{tab:HgIS} (online material). Otherwise, lines are synthesized without this
data. This leads us to ignore lines with a large reduced $\chi^{2}$,
related to the difference between the observed and synthesized
lines. The same procedure is used for all analyzed
elements. Fig.~\ref{fit:MnHFS} shows the
Mn~{\scriptsize{II}}~$\lambda~4206$ line with the hfs data of the line
taken into account and without it. These results show that the
synthetic Mn~{\scriptsize{II}}~$\lambda~4206$ line is better fitted
when the hfs data is taken into account, except for the fast rotating
star HD~221507. For this star, no significant difference is observed
between the fits with and without hfs data. Nevertheless, the
abundance obtained from the hfs data of this line for HD~221507, is
closer to the mean abundance of Mn measured in the atmosphere of this
star. The consideration of the isotopic composition of
Hg~{\scriptsize{II}}~$\lambda~3984$ agrees  with observed lines (see
Fig.~\ref{fit:HgISS}). Both the core and wings of this line are
better fitted by taking into account the isotopic composition. The
isotopic mixture of mercury for the peculiar
Hg~{\scriptsize{II}}~$\lambda~3984$ line is presented in Table~\ref{tab:HgISC} (online material).
 
During the spectral synthesis, the continuous opacities for
H~{\scriptsize{I}}, H$^{-}$ and He~{\scriptsize{I}} bound-free (b-f)
and free-free (f-f) transitions are computed. Photons scattering by
free electrons and H~{\scriptsize{I}} atoms are also taken into
account in the opacity calculations, as well as line broadening due to
the natural, Doppler and pressure broadenings. For more details, see
Wade et al. (2001a). To fit a line, we simultaneously determine the
element's abundance log~($N_{ion}/N_{tot}$), the stellar radial
velocity $V_{\rm r}$ and projected rotational velocity $v~\sin~i$ by
minimizing the deviation between the simulated and observed line
profiles. The optical depth of formation of the line is defined as the
continuum optical depth $\tau_{5000}$ where the monochromatic optical
depth of the line centre is equal to unity (Khalack et al. 2007 and
2008).

\begin{table}
\center
\caption{Effective temperature, luminosity, mass and age of studied stars. The approximate errors in effective temperatures and luminosities are respectively $\pm 5\%$ and $\pm0.1$~dex.}
\label{tab:HR}
\begin{tabular}{ l c c c c c c  l c c c l l  }\hline\hline 
HD & $T_{\rm eff}$ & log~$(L/L_{\rm \odot}$) & M/M$_{\rm \odot}$  & log~$t$\\ 
          & (K) &  (dex)           &                  &  \\\hline 
71066  & 12 010 &  2.05  & 3.09   & 8.11   \\ 
175640 & 11 958 &  2.08  & 3.21   & 8.16  \\ 
178065 & 12 193 &  2.26  & 3.31   & 8.21  \\ 
221507 & 12 476 & 1.88  & 3.07   & 6.09  \\\hline
\end{tabular}
\end{table}

\begin{figure}
\includegraphics[scale=0.32,angle=-90]{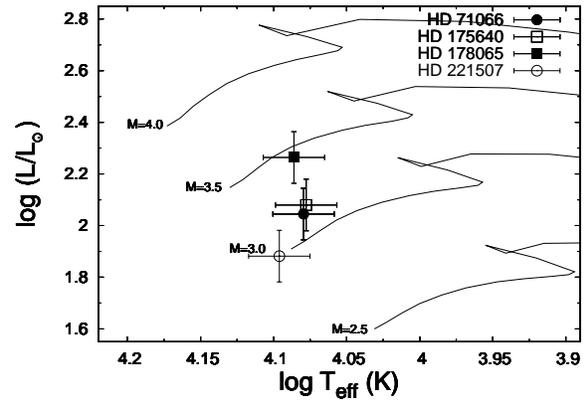}
\caption{The positions  of HD~71066 (filled  circle), HD~175640 (empty
square), HD~178065 (filled square) and HD~221507 (empty circle) in the Hertzsprung-Russell diagram  assuming   that  their  temperatures  and
luminosities  are  respectively  known   within  $\pm  5\%$  and  $\pm
0.1$~dex. These positions are  determined using the standard evolution
track Z=0.02 (Schaller et al. 1992).}
\label{fig:HR}
\end{figure}

Two stars studied here are SB1 and the others are non-binary (see
Table~\ref{tab:tousensemble}). Since the contribution of the companion
star for SB1 systems is not observed in the spectra, in this paper the
SB1 stars are treated as single stars. The position in the
Hertzsprung-Russell (HR) diagram of each star was determined
photometrically. The effective temperature and surface gravity used
are those from Table~\ref{tab:tousensemble}. The Johnson magnitude and
Str\"omgren indices in the $ubvy$ photometric system are extracted
from the General Catalogue of Photometric
Data\footnote{http://obswww.unige.ch/gcpd/gcpd.html.}. Stars' parallax
values are extracted from Hipparcos
catalogue\footnote{http://cadcwww.hia.nrc.ca/astrocat/hipparcos/}. The
bolometric correction was calculated as in Balona (1994). The
parameters such as log~($L/L_{\rm \odot}$), M/M$_{\rm \odot}$ and
log~$t$ presented in Table~\ref{tab:HR}, are determined
using the evolutionary model calculations of Schaller et al. (1992)
with ${\rm Z}=0.02$. The position of stars in the HR diagram
(log~($L/L_{\rm \odot}$), log~($T_{\rm eff}$)) is presented in
Fig.~\ref{fig:HR}. As shown in this figure, all stars studied here
belong to the main sequence. HD~221507 is the youngest star, while the
others appear to be somewhat older.

\section{Average abundances}

\begin{table*}
\begin{minipage}{140mm}
\caption{Mean  abundances log~($N_{ion}/N_{tot}$) of chemical species
measured  in the  atmospheres  of HD~71066,  HD~175640, HD~178065  and
HD~221507.}
\label{tab:average}
\begin{tabular}{l l l l l r c r}\hline\hline
Ion & HD 71066 & HD 175640 & HD 178065 & HD 221507 & \multicolumn{3}{c}{Sun}\\ & & & & & [A] & [B] & [C]\\\hline 
He {\scriptsize{I}} & $-2.30 \pm 0.40$ & $-1.65 \pm 0.13$ & $-1.78 \pm 0.12$ & $-1.93 \pm 0.14$ & $-1.05$ & $-1.07 \pm 0.01$ & $0.02$\\ 
C {\scriptsize{II}} & $-3.89 \pm 0.10$ & $-3.86 \pm 0.10^{*}$ & $-3.78 \pm 0.05^{*}$ & $-3.92 \pm 0.05^{*}$ & $-3.48$ & $-3.61 \pm 0.05$ & $0.13$\\ 
O {\scriptsize{I}} & $-3.61 \pm 0.14$ & $-3.24 \pm 0.04$ & $-3.33 \pm 0.15$ & $-3.11 \pm 0.09^{*}$ & $-3.11$ & $-3.34 \pm 0.05$ & $0.23$ \\ 
Mg {\scriptsize{I}} & ...  & $-4.78 \pm 0.06^{*}$ & $-4.86 \pm 0.04^{*}$ & $-4.40 \pm 0.10^{*}$ & $-4.46$ & $-4.47 \pm 0.09$ & $0.01$ \\ 
Mg {\scriptsize{II}} & $-5.46 \pm 0.01$ & $-4.86 \pm 0.01$ & $-5.02 \pm 0.03$ & $-4.73 \pm 0.18$ & $-4.46$ & $-4.47 \pm 0.09$ & $0.01$ \\ 
Si {\scriptsize{II}} & $-4.58 \pm 0.07$ & $-4.69 \pm 0.03$ & $-4.58 \pm 0.12$ & $-4.53 \pm 0.06$ & $-4.49$ & $-4.49 \pm 0.04$ & $0.00$ \\ 
P {\scriptsize{II}} & $-4.87 \pm 0.22$ & $-6.37 \pm 0.15$ & $-5.26 \pm 0.15$ & $-5.42 \pm 0.38$ & $-6.59$ & $-6.64 \pm 0.04$ & $0.05$ \\ 
S {\scriptsize{II}} & $-5.66 \pm 0.20$ & $-5.05 \pm 0.16$ & $-5.62 \pm 0.13$ & $-5.13 \pm 0.32$ & $-4.83$ & $-4.86 \pm 0.05$ & $0.03$ \\
Ca {\scriptsize{II}} & $-6.02 \pm 0.04^{*}$ & $-5.38 \pm 0.14$ & $-5.87 \pm 0.04^{*}$ & $-5.07 \pm 0.03^{*}$ & $-5.68$ & $-5.69 \pm 0.04$ & $0.01$\\ 
Sc {\scriptsize{II}} & ...  & ...  & $-8.68 \pm 0.04$ & ...  & $-8.94$ & $-8.83 \pm 0.10$ & $-0.11$\\ 
Ti {\scriptsize{II}} & $-6.52 \pm 0.05$ & $-5.87 \pm 0.07$ & $-6.58 \pm 0.04$ & $-6.55 \pm 0.13$ & $-7.05$ & $-7.10 \pm 0.06$ & $0.05$\\ 
Cr {\scriptsize{I}} & ...  & $-5.50 \pm 0.03$ & ...  & ...  & $-6.37$ & $-6.36 \pm 0.10$ & $-0.01$\\ 
Cr {\scriptsize{II}} & $-6.28 \pm 0.09$ & $-5.55 \pm 0.18$ & $-6.08 \pm 0.11$ & $-5.92 \pm 0.12$ & $-6.37$ & $-6.36 \pm 0.10$ & $-0.01$\\ 
Mn {\scriptsize{I}} & ...               & $-4.44 \pm 0.10$ & $-5.07 \pm 0.16$ & $-4.38 \pm 0.21$ & $-6.65$ & $-6.61 \pm 0.03$ & $-0.04$\\ 
Mn {\scriptsize{II}} & $-5.81 \pm 0.20$ & $-4.51 \pm 0.18$ & $-5.08 \pm 0.24$ & $-4.36 \pm 0.22$ & $-6.65$ & $-6.61 \pm 0.03$ & $-0.04$\\ 
Fe {\scriptsize{I}} & $-3.98 \pm 0.06$ & $-4.91 \pm 0.04$ & $-4.60 \pm 0.09$ & $-4.33 \pm 0.13$ & $-4.37$ & $-4.55 \pm 0.05$ & $0.18$\\ 
Fe {\scriptsize{II}} & $-3.87 \pm 0.14$ & $-4.88 \pm 0.12$ & $-4.57 \pm 0.09$ & $-4.41 \pm 0.17$ & $-4.37$ & $-4.55 \pm 0.05$ & $0.18$\\ 
Fe {\scriptsize{III}} & $-3.84 \pm 0.07$ & ...  & ...  & ...  & $-4.37$ & $-4.55 \pm 0.05$ & $0.18$\\ 
Ni {\scriptsize{II}} & ...  & $-6.46 \pm 0.04^{*}$ & $-6.60 \pm 0.01$ &$-6.14 \pm 0.07^{*}$ & $-5.79$ & $-5.77 \pm 0.04$ & $-0.02$\\ 
Sr {\scriptsize{II}} & $-8.35 \pm 0.03^{*}$ & $-8.61 \pm 0.02^{*}$ & $-9.21 \pm 0.01^{*}$ & $-6.00 \pm 0.04^{*}$ & $-9.14$ & $-9.08 \pm 0.05$ & $-0.06$\\ 
Hg {\scriptsize{I}} & $-6.38 \pm 0.28$ & $-6.13 \pm 0.20^{*}$ & $-6.36 \pm 0.19^{*}$ & $-4.59 \pm 0.35^{*}$ & $-10.95$ & $-10.87 \pm 0.18^{**}$ & $-0.08$\\ 
Hg {\scriptsize{II}} & $-6.53 \pm 0.33$ & $-6.38 \pm 0.20^{*}$ & $-6.89 \pm 0.19$ & $-5.61 \pm 0.13$ & $-10.95$ &$-10.87 \pm 0.18^{**}$ & $-0.08$\\\hline
\end{tabular}
Elements for which abundances are derived from a unique line are marked with an asterisk (*).

[A] and [B] are respectively the chemical solar abundances used in
Kurucz (1993) and Grevesse et al. (2007). Hg abundance of [B] marked
with double asterisks (**) is obtained from meteorites. The abundance
difference between these two sets are listed in column [C].
\end{minipage}
\end{table*}

The mean abundances log~($N_{ion}/N_{tot}$) measured in the
atmospheres of the HgMn stars HD~71066, HD~175640, HD~178065 and
HD~221507 are reported in Table~\ref{tab:average}. Several error
sources have to be taken into account. For example, Khan \& Shulyak
(2007) show that by using an atmospheric model calculated with iron
abundance [Fe/H] between 1 and 10 times its solar value, the derived
abundances for elements can change by up to $\pm
0.25$~dex. Uncertainties in atmospheric parameters ($T_{\rm eff}$ and
log~$g$) and in the atomic data can also modify abundances. NLTE
effects may also play a role. The errors in the average abundances in
Table~\ref{tab:average} are obtained by calculating the standard
deviation relative to the mean abundance for elements with a large
number of lines. The uncertainties in the atmospheric parameters are
not included in error calculations for these elements. For those
species for which abundances are derived from a unique line, we took
the formal error derived from the line profile fit. This error arises
from the uncertainties in stellar atmospheric parameters and atomic
data.  The solar abundances used in this study are those used in
Kurucz (1993)(see [A] of Table~\ref{tab:average}).  In the same table,
we present the solar abundances determined by Grevesse et al. (2007)
([B]).  The differences between these two sets of abundances are
summarized in column 8 ([C]) of Table~\ref{tab:average} and are within
$\sim 0.2$~dex.

\begin{table*}
\begin{minipage}{126mm}
\caption{The slope $a$, obtained from a linear regression of abundance
vs. optical depth: ${\rm log}~(N_{ion}/N_{tot})~=~a~{\rm
log}~\tau_{5000}~+~b$. The number of lines investigated is represented
by $n$.}
\label{tab:LSM}
\begin{tabular}{ c c c c c c c c  c }\hline\hline
Element & \multicolumn{2}{c}{HD~71066}  & \multicolumn{2}{c}{HD~175640} & \multicolumn{2}{c}{HD~178065}
        & \multicolumn{2}{c}{HD~221507} \\ 
& $a$                & $n$ & $a$                & $n$      & $a$                & $n$ & $a$                & $n$      \\\hline 
S       & ...                & ...      & $+0.043 \pm 0.1761$ & $15$     & ... & ...      & ...                & ...      \\ 
Ti      & $+0.050 \pm 0.061$ & $28$     & $-0.069 \pm 0.025$ & $41$     & $+0.034 \pm 0.021$ & $33$     & $+0.130 \pm 0.143$ & $14$ \\ 
Cr      & $+0.045 \pm 0.065$ & $19$     & $-0.010 \pm 0.046$ & $45$     & $-0.025 \pm 0.039$ & $35$     & $+0.072 \pm 0.079$ & $12$      \\ 
Mn      & ...                & ...      & $+0.008 \pm 0.038$ & $62$     & $+0.289 \pm 0.060$ & $62$     & $+0.051 \pm 0.069$ & $28$     \\ 
Fe      & $+0.035 \pm 0.012$ & $158$    & $+0.061 \pm 0.018$ & $57$     & $+0.052 \pm 0.013$ & $94$     & $+0.008 \pm 0.024$ & $57$     \\\hline
\end{tabular}
\end{minipage}
\end{table*}

\subsection{HD~71066}

The elements He, C, O, Mg, Si, P, S, Ca, Ti, Cr, Mn, Fe, Sr and Hg
were investigated to determine their abundances in the atmosphere of
HD~71066. The abundances are given in col.~2 of
Table~\ref{tab:average}. Comparing these abundances to those of the
sun, the elements C, O, Si, Ca, Ti, Cr and Fe are within 0.5~dex of
solar values.  He, Mg and S are deficient by more than 0.5~dex, while
P, Mn, Sr and Hg are overabundant by more than 0.5~dex.

Comparing the Cr abundance derived in this paper to the
log~($N_{Cr}/N_{H}$) $=-6.23\pm0.10$ obtained by Savanov \& Hubrig
(2003), we find a good agreement. The enhanced Fe abundance of
0.11~dex obtained by Hubrig et al. (1999) is consistent with our
measurement when taking into account the error bars. Concerning the
abundance of Hg, no significant difference is observed between that
derived from this study and the value reported by Dolk et al. (2003),
namely log~($N_{Hg}/N_{tot}$) $=-6.35\pm0.26$.

\subsection{HD~175640}

Chemical abundances of a large number of elements were determined for
HD~175640 by Castelli \& Hubrig (2004). In this paper, we diagnosed
abundances of He, C, O, Mg, Si, P, S, Ca, Ti, Cr, Mn, Fe, Ni, Sr and
Hg. The measured abundances are presented in col.~3 of
Table~\ref{tab:average}.  The elements C, O, Mg, Si, P, S and Ca show abundances within 0.5~dex of solar values. Helium, iron and nickel are
underabundant by more than 0.5 dex, while Ti, Cr, Mn, Sr and Hg are
enhanced by more than 0.5~dex.

The abundances obtained in this study are generally in agreement with
those derived by Castelli \& Hubrig (2004). However, our abundance of
Ni is 0.37~dex lower than reported by those authors
(log~($N_{Ni}/N_{tot}$) $=-6.09\pm0.16$). The mercury abundance
log~($N_{Hg}/N_{tot}$) $=-6.35\pm0.15$ derived by Dolk et al. (2003)
is close to that reported in this paper. Disagreements are observed between mean abundances derived from the analysis
of neutral and singly ionized Hg (0.25~dex) lines. These differences are greater than
the error bars associated with these abundances. Potential sources of
these differences in the abundances inferred from different ions of an
element, which are observed in several of the stars in this study,
will be explored in Sect. 6.

Unlike the other stars presented in this paper, the spectrum of
HD~175640 does not show any absorption (or emission) line at the
location of the Hg~{\scriptsize{II}} $\lambda 5677$ line.

\subsection{HD~178065}

Lines of the elements He, C, O, Mg, Si, P, S, Ca, Sc, Ti, Cr, Mn, Fe,
Ni, Sr and Hg were analyzed to determine their abundances in the
atmosphere of HD~178065.  The abundances of the elements C, O, Mg, Si,
Ca, Sc, Ti, Cr, Fe and Sr are within 0.5~dex of their corresponding
solar values. He, S and Ni are deficient by more than 0.5~dex, while
P, Mn and Hg are enhanced by more than 0.5~dex.

The Cr abundance is close to the log~($N_{Cr}/N_{H}$) $=-5.99\pm0.10$
obtained by Savanov \& Hubrig (2003). A slight difference of 0.14~dex
is obtained between the iron abundance derived in this study and the
log~($N_{Fe}/N_{tot})=-4.43$ obtained by Hubrig \& Castelli
(2001). The mean abundance of Hg is in good agreement with
log~($N_{Hg}/N_{tot})=-6.65\pm0.12$ derived by Dolk et al. (2003).

As mentioned above, differences of 0.16~dex and 0.53~dex are
respectively observed between the abundances of neutral and
singly-ionized states of both Mg and Hg.

\subsection{HD~221507}

For the more rapidly rotating ($v~\sin~i=25$~km s$^{-1}$) star
HD~221507, we analyzed abundances of He, C, O, Mg, Si, P, S, Ca, Ti,
Cr, Mn, Fe, Ni, Sr and Hg. We find that C, O, Mg, Si, S, Ti, Cr, Fe and Ni are within 0.5~dex of the solar abundance. Helium is the only underabundant element, while P, Ca, Mn, Sr and Hg are overabundant by more than 0.5~dex.

As observed in other stars, a difference (of 0.33~dex) is observed
between abundances of neutral and singly ionized magnesium. The
abundance derived from the Hg~{\scriptsize{I}} $\lambda5460.731$ line
is enhanced by 1.02~dex with respect to that obtained from
Hg~{\scriptsize{II}} lines.

The Si abundance agrees very well with log~($N_{Si}/N_{tot}$)$=-4.5\pm0.10$ obtained from UV analysis by Smith (1993). Our result for Mn abundance is consistent with the value (log~($N_{Mn}/N_{H}$)$=-4.5\pm0.05$) found by Jomaron et al. (1999) for this star. The iron abundance is in excellent agreement with log~($N_{Fe}/N_{H}$)$=-4.35\pm0.10$ obtained by Smith \& Dworetsky (1993). Although the Hg~{\scriptsize{II}} abundance obtained by Dolk et al. (2003) differs by $0.21\pm0.43$~dex compared to our analysis, the values are still within the error bars.

\section{Stratification}

For elements exhibiting a sufficiently large number of suitable lines
in our spectra, we have examined the dependence of their abundance
versus the optical depth of line core formation. To evaluate vertical
stratification, we have calculated the slope $a$ of the abundance
versus $\tau_{5000}$ by performing a linear fit to the data by the
Least Squares Method (LSM): ${\rm log}~(N_{ion}/N_{tot})~=~a~{\rm
log}~\tau_{5000}~+~b$. The slopes $a$ obtained for investigated
elements are summarized in Table~\ref{tab:LSM}. For the confirmation
of vertical stratification, we require that the three following
conditions be met:

\begin{enumerate}
\renewcommand{\theenumi}{(\arabic{enumi})}
\item More than 10 lines, formed at different optical depths, must be
present in the spectra.
\item The slope $a$ should be statistically significant in comparison
to the uncertainties.
\item The chemical abundance should change significantly (by more than
$\sim$0.5~dex) in the diagnosed range of optical depth.
\end{enumerate}

This last condition is related to the fact that several factors such
as errors in atmospheric parameters ($T_{\rm eff}$ and log~$g$), NLTE
effects and uncertainties in the model atmosphere can cause errors in
abundance determination. For example, as discussed by Takada-Hidai
(1991), an error of $\pm1000$~K in $T_{\rm eff}$ causes errors in
abundances derived from resonance lines by $\pm0.08$~dex and
$[-0.19,+0.25]$~dex for lines with high excitation
energy. Furthermore, an error of $\pm0.3$~dex in log~$g$, introduces
uncertainties in abundance calculations of $\pm0.1$~dex. The
uncertainty in the metallicity can account for an error of
$\sim0.25$~dex for abundance calculations (Khan \& Shulyak
2007). Requiring a significant change in abundance helps ensure that
any abundance gradient is real, and not generated by poorly-quantified
uncertainties.

\begin{figure}
\center
\vspace{174pt}
\includegraphics[scale=0.33,angle=-90]{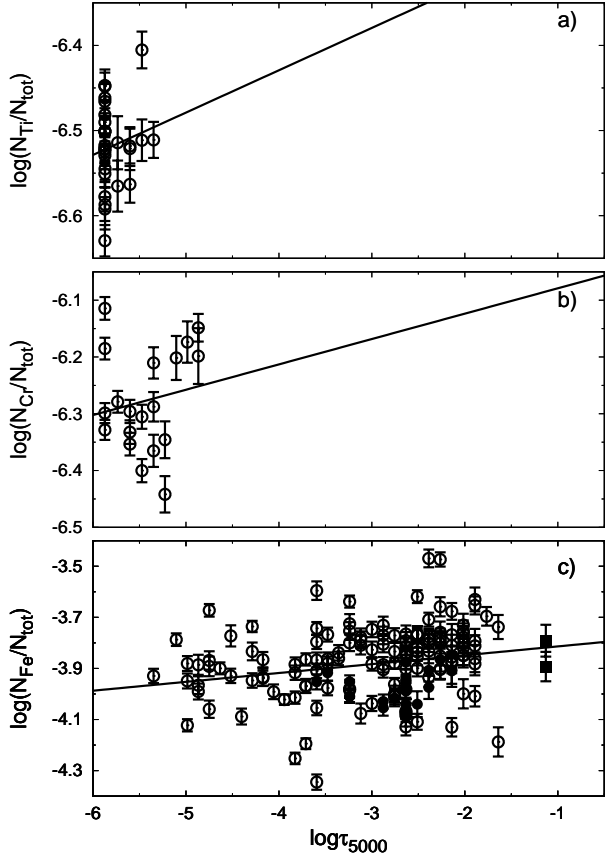}
\vspace{-20pt}
\caption{Dependence of a) Ti, b) Cr and c) Fe abundance in the
atmosphere of HD~71066 against the line (core) formation optical depth
calculated at $\lambda = 5000$~\AA. The optical depth range between
$-6$ and $-0.5$ is approximatively associated to the physical depth
from $\sim0$ to $9.58\times10^{3}$~km. The filled circles, open
circles and filled squares correspond respectively to abundances
derived from the neutral, singly and doubly ionized ions. The error
bars correspond to formal uncertainties derived from the fits of line
profiles.  The linear fit of the data using the least squares method
${\rm log}~(N_{ion}/N_{tot})~=~a~{\rm log}~\tau_{5000}~+~b$ is
represented by the solid line.}
\label{fig:tau_hd71066}
\end{figure}

\begin{figure}
\center
\vspace{378pt}
\includegraphics[scale=0.33,angle=-90]{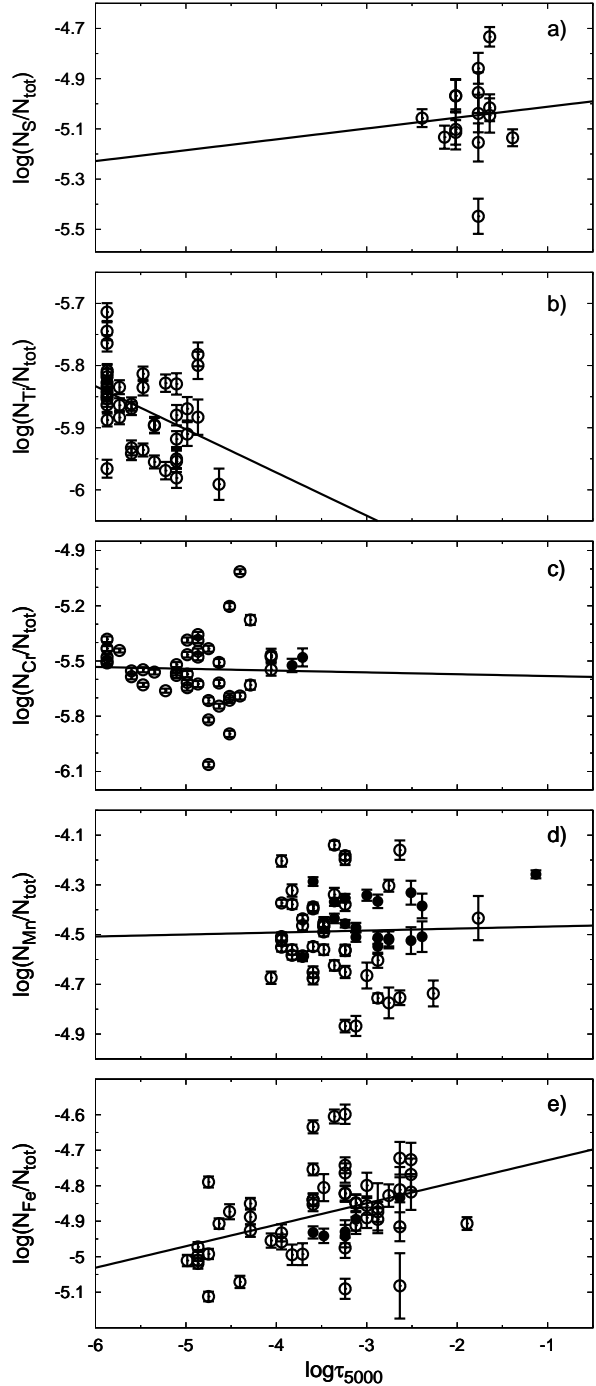}
\vspace{-20pt}
\caption{Same description as in Fig.~\ref{fig:tau_hd71066} but for a)
S, b) Ti, c) Cr, d) Mn and e) Fe for HD~175640.}
\label{fig:tau_hd175640}
\end{figure}

\subsection{HD~71066}

In the spectrum of HD~71066, lines of Ti, Cr and Fe were studied for
vertical stratification. We found that titanium lines are formed over
a small range of optical depth, between log~$\tau_{5000} \approx -5.9$
and $-5.3$ (see Fig.~$\ref{fig:tau_hd71066}a$). The slope $a$ is
relatively small compared to the formal uncertainty (see
Table~\ref{tab:LSM}). We conclude that the titanium abundance does not
show a significant dependence on optical depth for HD~71066.

As reported in Table~\ref{tab:LSM}, the linear fit of chromium
abundance versus optical depth reveals a slope smaller than the
error. No significant variation of Cr abundance in the optical depth
range from log~$\tau_{5000} \approx -5.9$ to $-4.8$ is detected (see
Fig.~$\ref{fig:tau_hd71066}b$).

From log~$\tau_{5000} \approx -5.5$ to $-1$, the iron abundance
(derived from 22 Fe~{\scriptsize{I}}, 134 Fe~{\scriptsize{II}} and 2
Fe~{\scriptsize{III}} lines) does not show a clear dependence on depth
(see Fig.~$\ref{fig:tau_hd71066}c$). The LSM gives a slope which is
marginally significant compared to the error ($a \approx 3
\sigma_{a}$). However, the variation of the abundance versus the
optical depth is too weak to confidently conclude that Fe is
stratified in the atmosphere of HD~71066.

\subsection{HD~175640}

In the spectrum of HD~175640, lines of S, Ti, Cr, Mn and Fe were
studied for vertical stratification. With 15 S~{\scriptsize{II}}
lines, we were able to evaluate the possibility of stratification of
this element in the atmosphere of HD~175640. As the slope $a$ is lower
than the corresponding error, sulfur shows no sign of stratification
between log~$\tau_{5000} \approx -2.4$ and $-1.2$ (see
Fig.~$\ref{fig:tau_hd175640}a$).

The linear fit of Ti abundance versus the optical depth of HD~175640
reveals a weak (negative) slope. No significant variation of Ti
abundance is observed in the optical depth range from log~$\tau_{5000}
\approx -5.9$ to $-4.6$ of HD~175640 in Fig.~$\ref{fig:tau_hd175640}b$.

Forty-five Cr lines were studied in the spectrum of HD~175640, of
which 2 are from the neutral ion. The chromium abundance presented in
Fig.~$\ref{fig:tau_hd175640}c$ does not show any variation with
optical depth ranging between log~$\tau_{5000} \approx -5.9$ and $-3.6$.
The slope $a \approx 0.2 \sigma_{a}$ is not statistically significant
and is very weak. No stratification is thus observed.

Fig.~$\ref{fig:tau_hd175640}d$ shows the variation of Mn abundance
with optical depth between log~$\tau_{5000} \approx -4.1$ and
$-1$. The slope obtained from the linear fit of Mn abundance in the
atmosphere of HD~175640 is relatively weak and is statistically not significant ($a \approx 0.2 \sigma_{a}$). Manganese is not stratified in the atmosphere of HD~175640.

The investigation of 6 Fe~{\scriptsize{I}} and 51 Fe~{\scriptsize{II}}
lines yields a weak slope (see Table~\ref{tab:LSM}) between
log~$\tau_{5000} \approx -5.$ and $-1.8$. We conclude that iron is
not detectably stratified in the diagnosed optical depths of the
atmosphere of HD~175640 (see Fig.~$\ref{fig:tau_hd175640}e$).

\subsection{HD~178065}

\begin{figure}
\center
\vspace{275pt}
\includegraphics[scale=0.33,angle=-90]{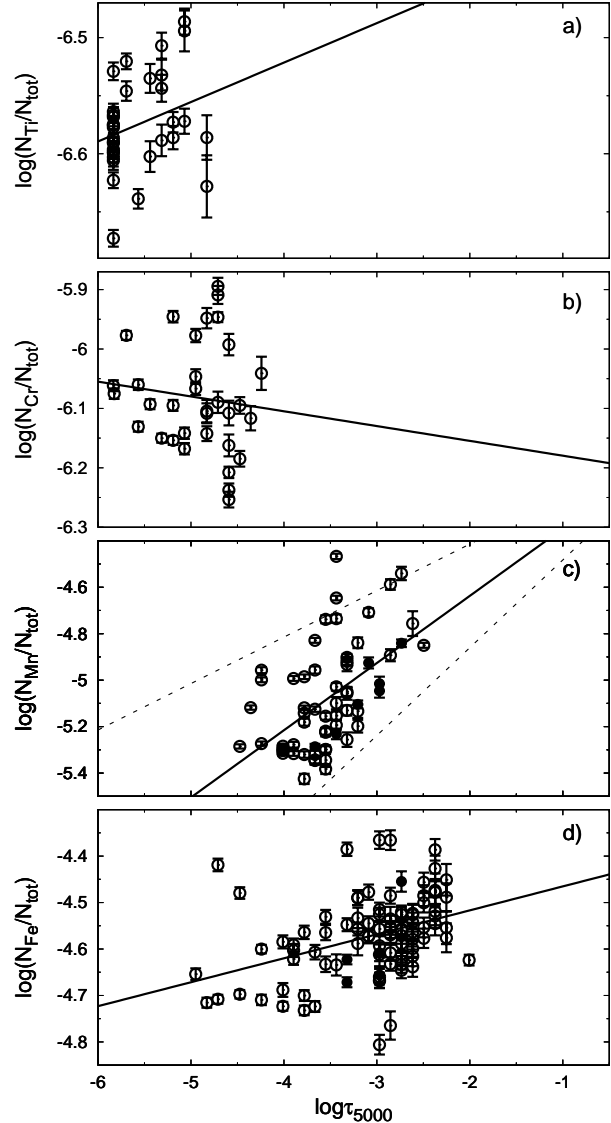}
\vspace{-20pt}
\caption{Same description as in Fig.~\ref{fig:tau_hd71066} but for a)
Ti, b) Cr, c) Mn and d) Fe for HD~178065, but where we have added
$3\sigma$ regression curves (dash) for Mn.}
\label{fig:tau_hd178065}
\end{figure}

In the spectrum of HD~178065, lines of Ti, Cr, Mn and Fe were studied
for vertical stratification. Fig.~$\ref{fig:tau_hd178065}a$ shows that
the titanium lines observed are formed in a small range of optical
depth (between log~$\tau_{5000} \approx -5.9$ and $-4.8$). The slight
slope (see Table~\ref{tab:LSM}) leads us to conclude that Ti is not
significantly stratified in the diagnosed optical depths of HD~178065.

As presented in Table~\ref{tab:LSM}, the linear fit of Cr abundance in
the investigated optical depth of HD~178065 reveals a negative
slope. However, this slope is weaker than the associated error ($a
\approx 0.6 \sigma_{a}$). The chromium abundance does not show a
systematic dependence on optical depth from log~$\tau_{5000} \approx
-5.9$ to $-4.2$ (see Fig.~$\ref{fig:tau_hd178065}b$).

The abundance obtained from 5 Fe~{\scriptsize{I}} and 89
Fe~{\scriptsize{II}} lines also does not reveal a significant
variation at optical depths between log~$\tau_{5000} \approx -5$ and
$-2$ (see Fig.~$\ref{fig:tau_hd178065}d$). The slope $a$ is relatively
small (see Table~\ref{tab:LSM}), and we conclude that iron is not
detectably stratified in the atmosphere of HD~178065 in the diagnosed
optical depths.

In contrast to the behaviour observed for Ti, Cr and Fe,
Fig.~$\ref{fig:tau_hd178065}c$ shows a strong increase of the Mn
abundance with optical depth. The slope $a \approx 5 \sigma_{a}$ is
statistically significant, and the inferred Mn abundance increases by
$\sim0.7$~dex in the diagnosed range of optical depth between
log~$\tau_{5000} \approx -4.5$ and $-2.5$. Therefore Mn satisfies all
3 of our criteria required for the detection of stratification.

\begin{figure*}
\includegraphics[scale=0.22,angle=-90]{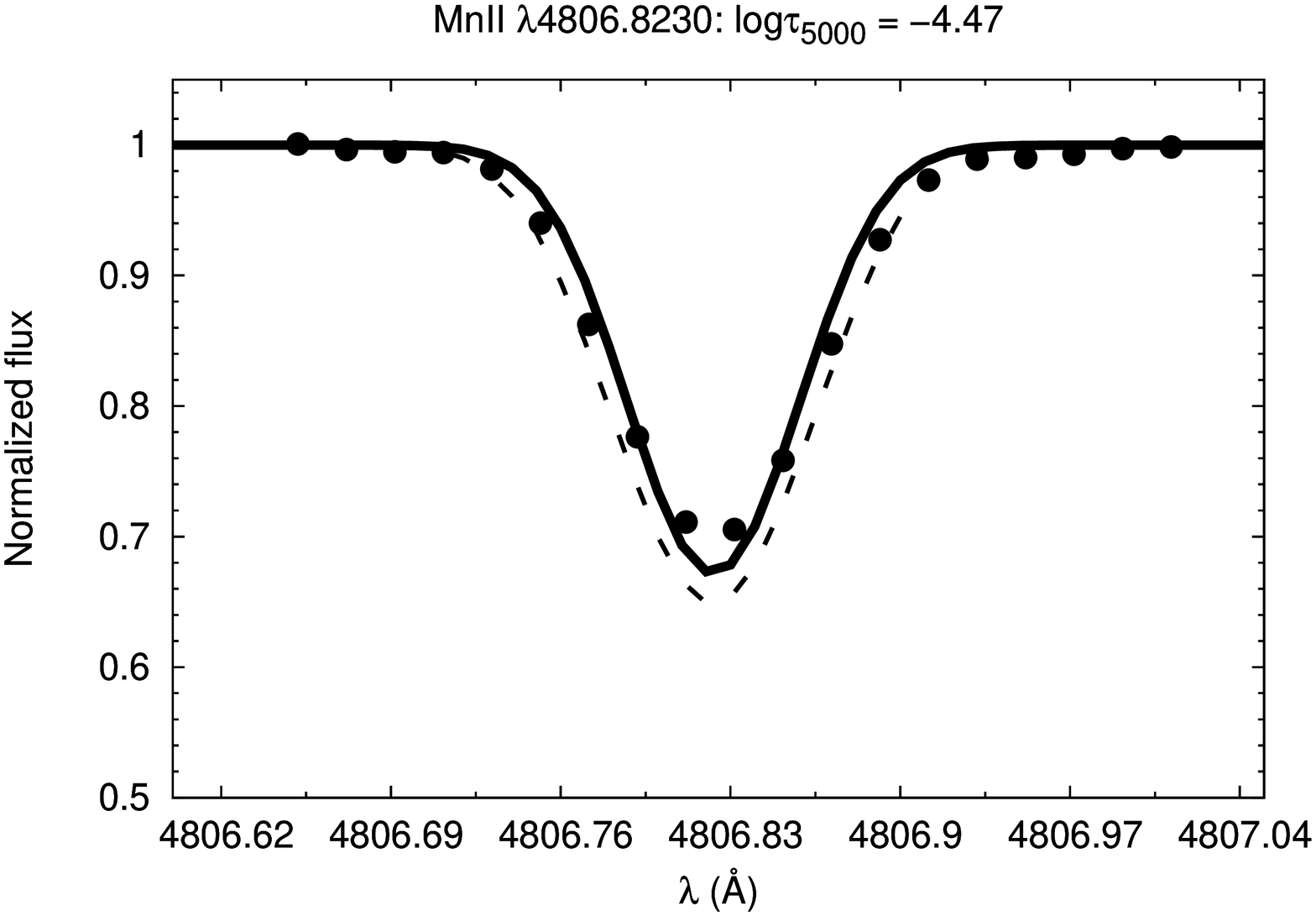}
\includegraphics[scale=0.22,angle=-90]{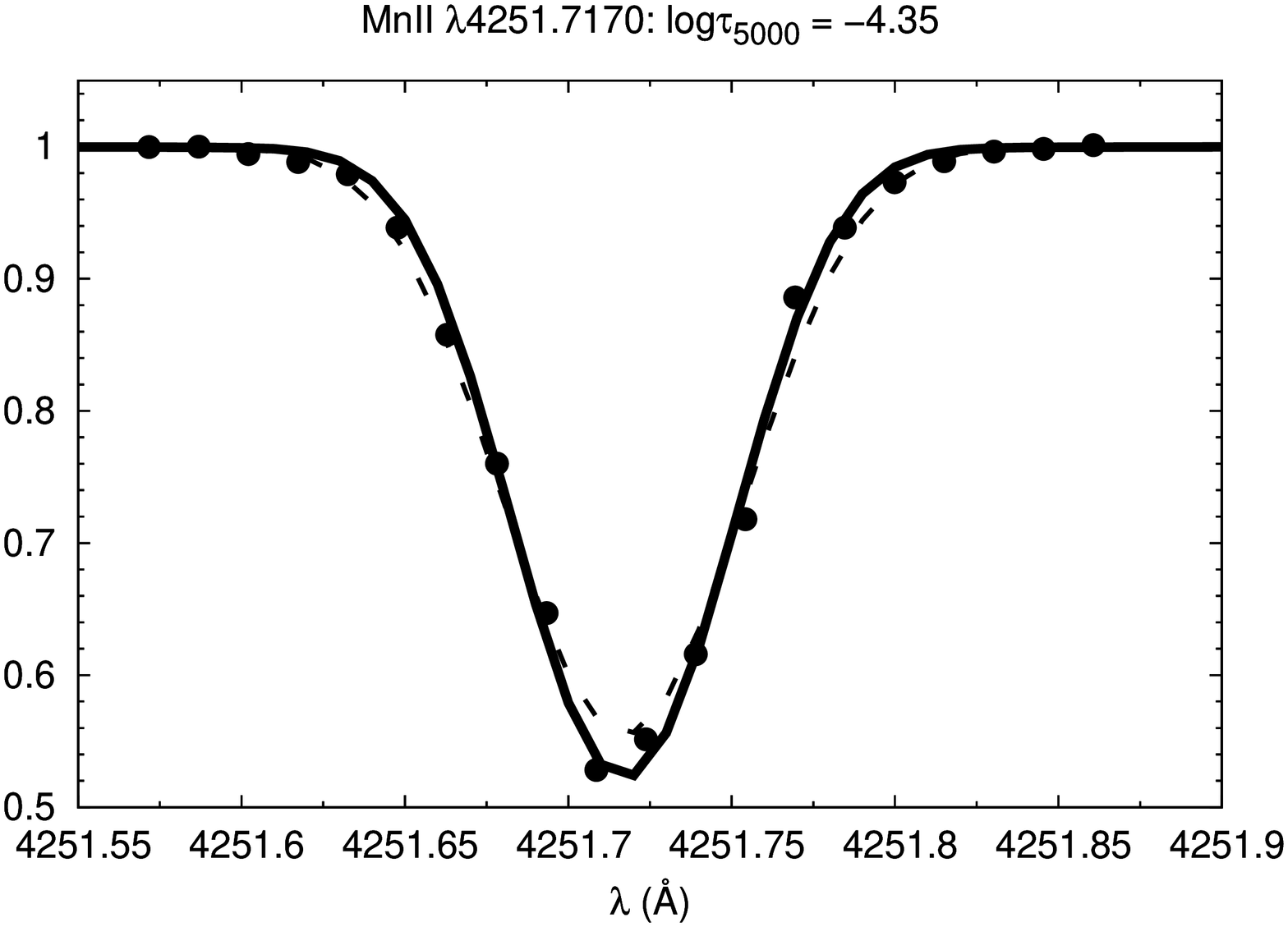}
\includegraphics[scale=0.22,angle=-90]{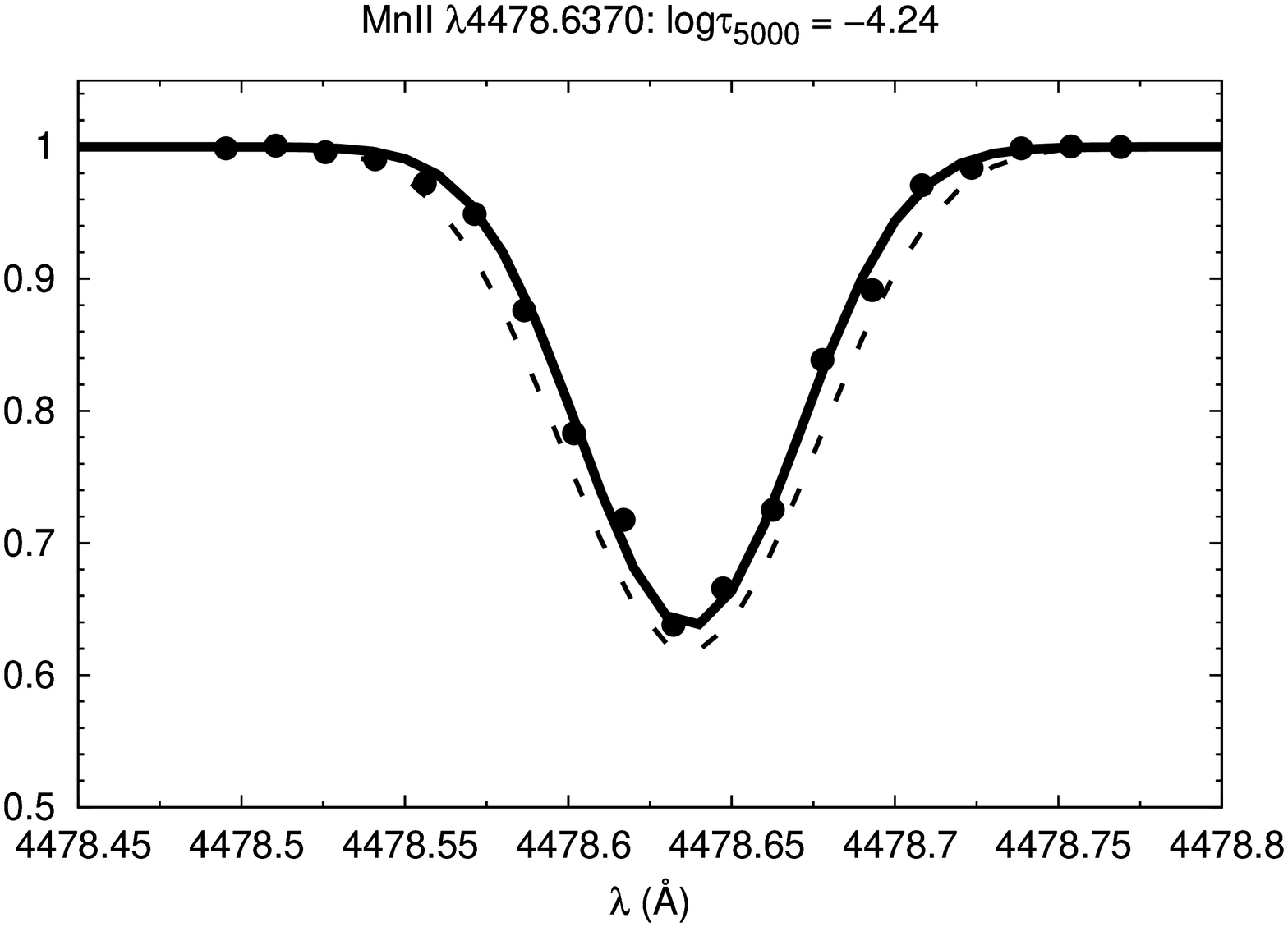}\\
\includegraphics[scale=0.22,angle=-90]{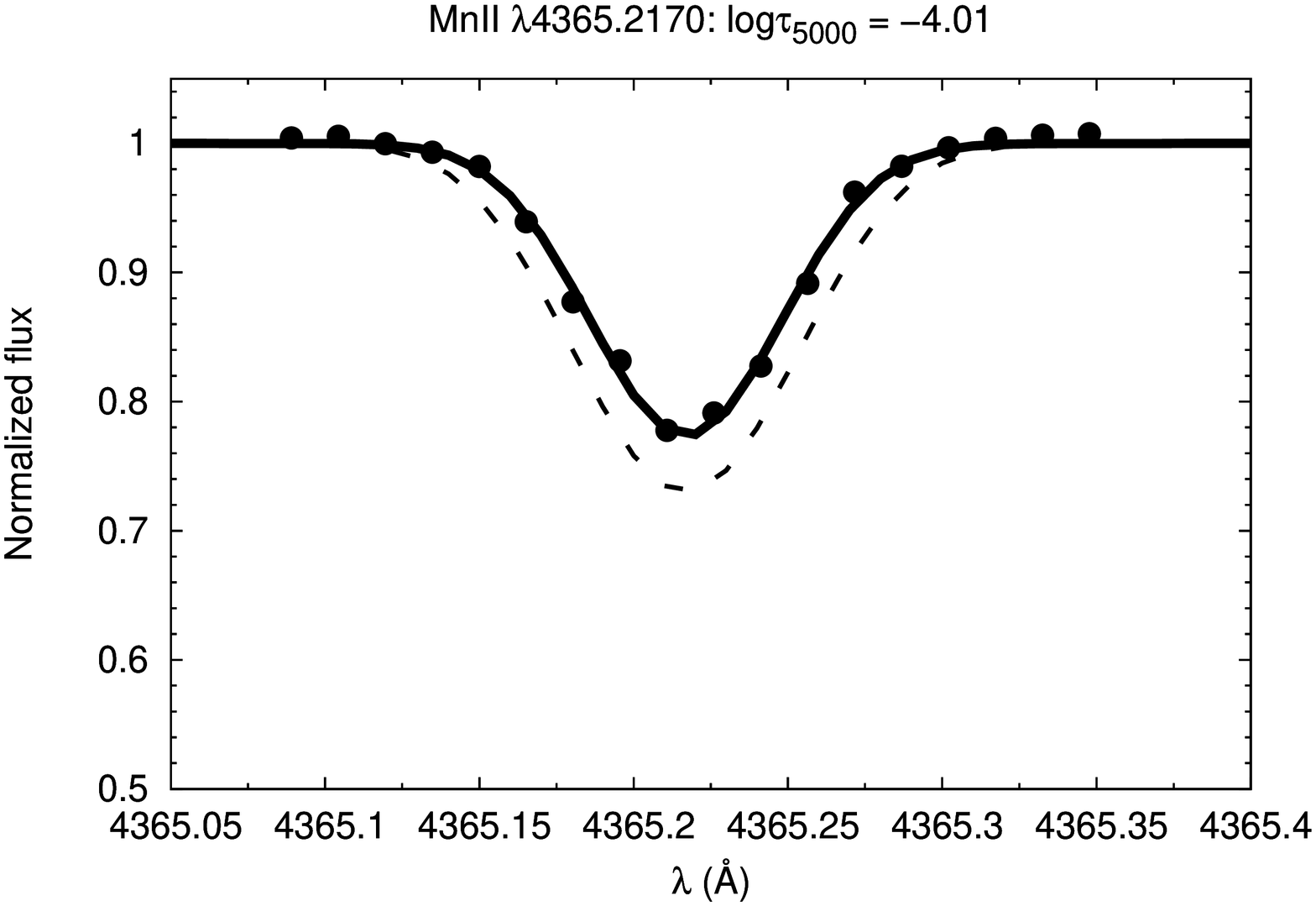}
\includegraphics[scale=0.22,angle=-90]{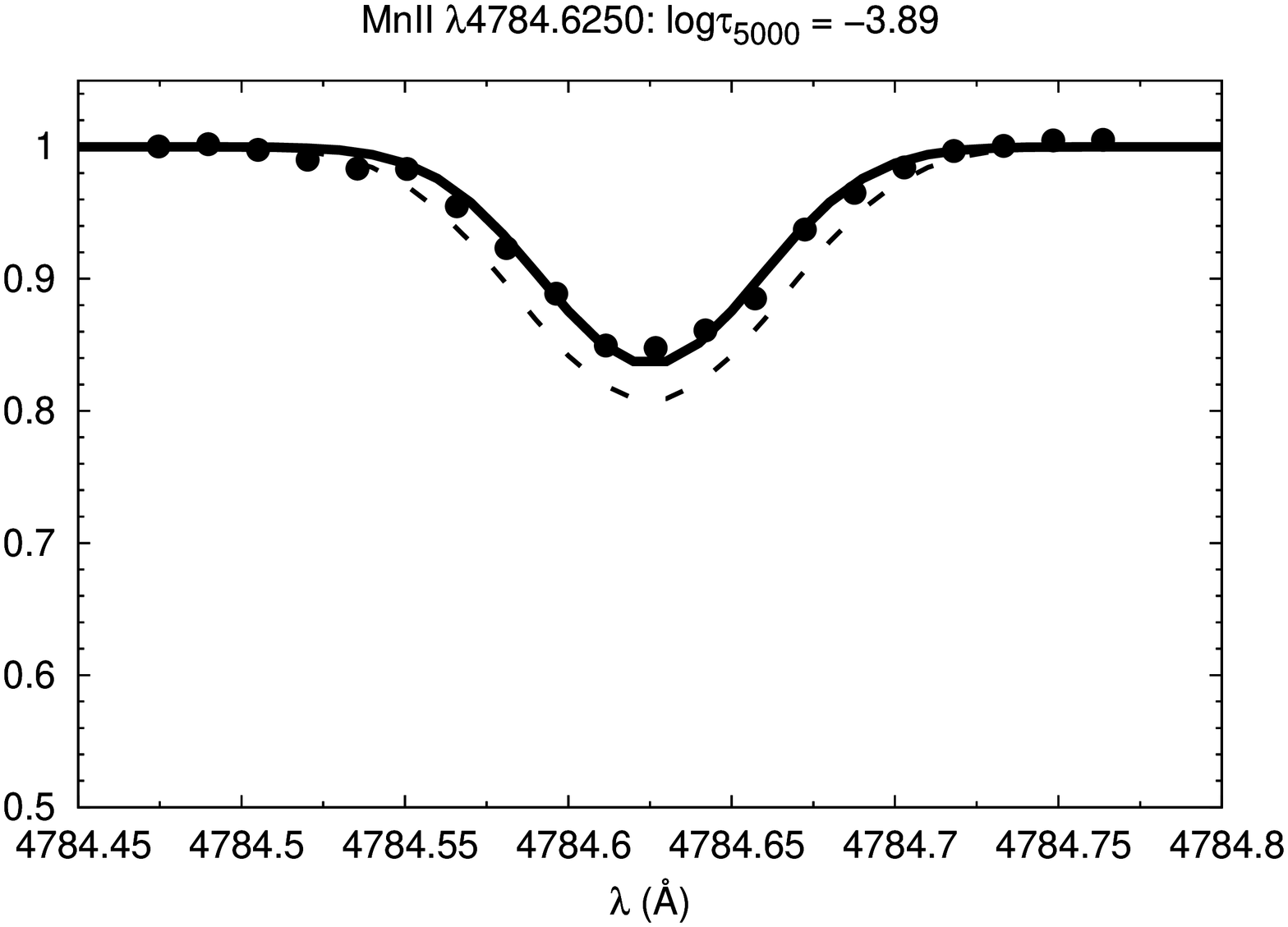}
\includegraphics[scale=0.22,angle=-90]{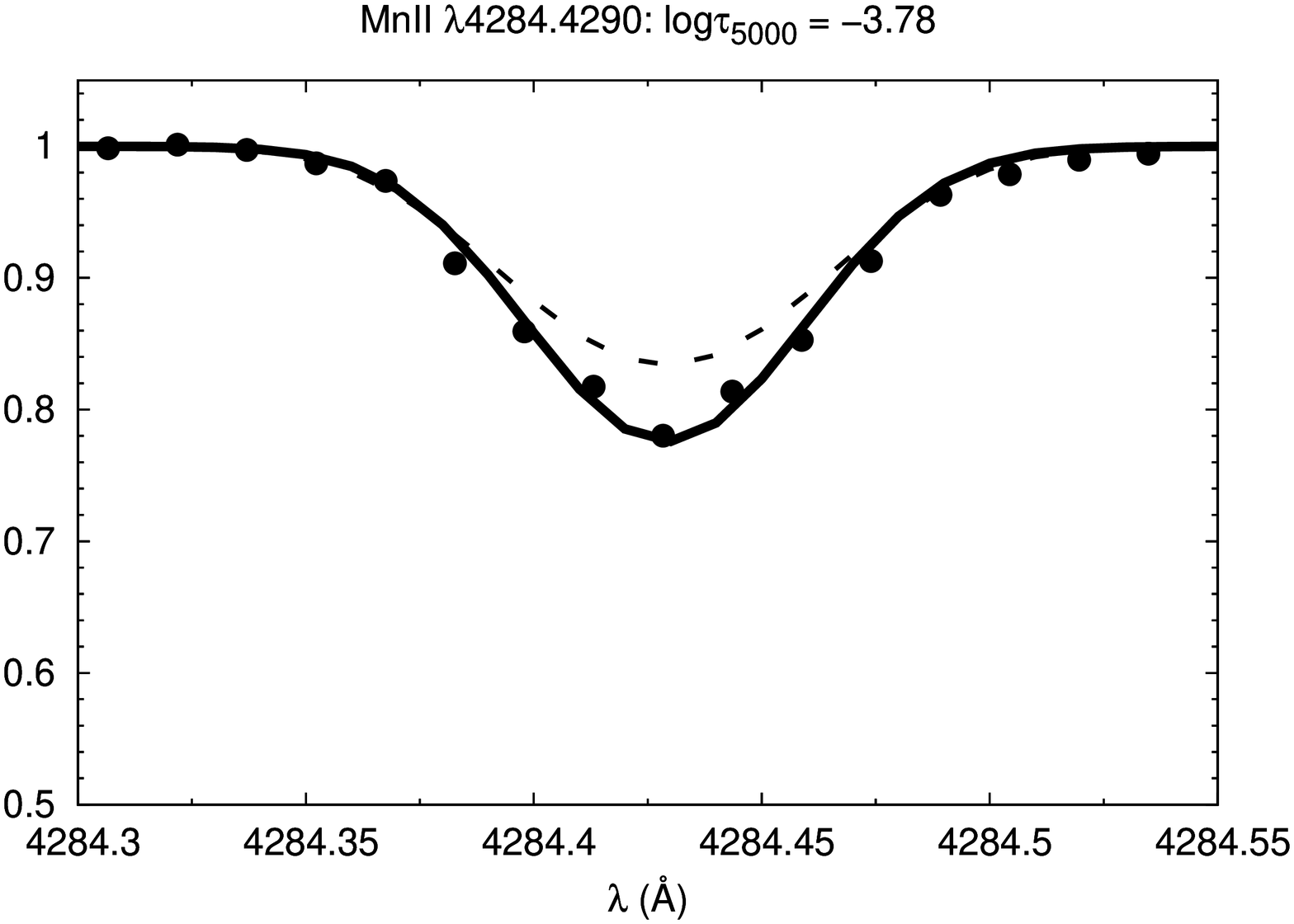}\\
\includegraphics[scale=0.22,angle=-90]{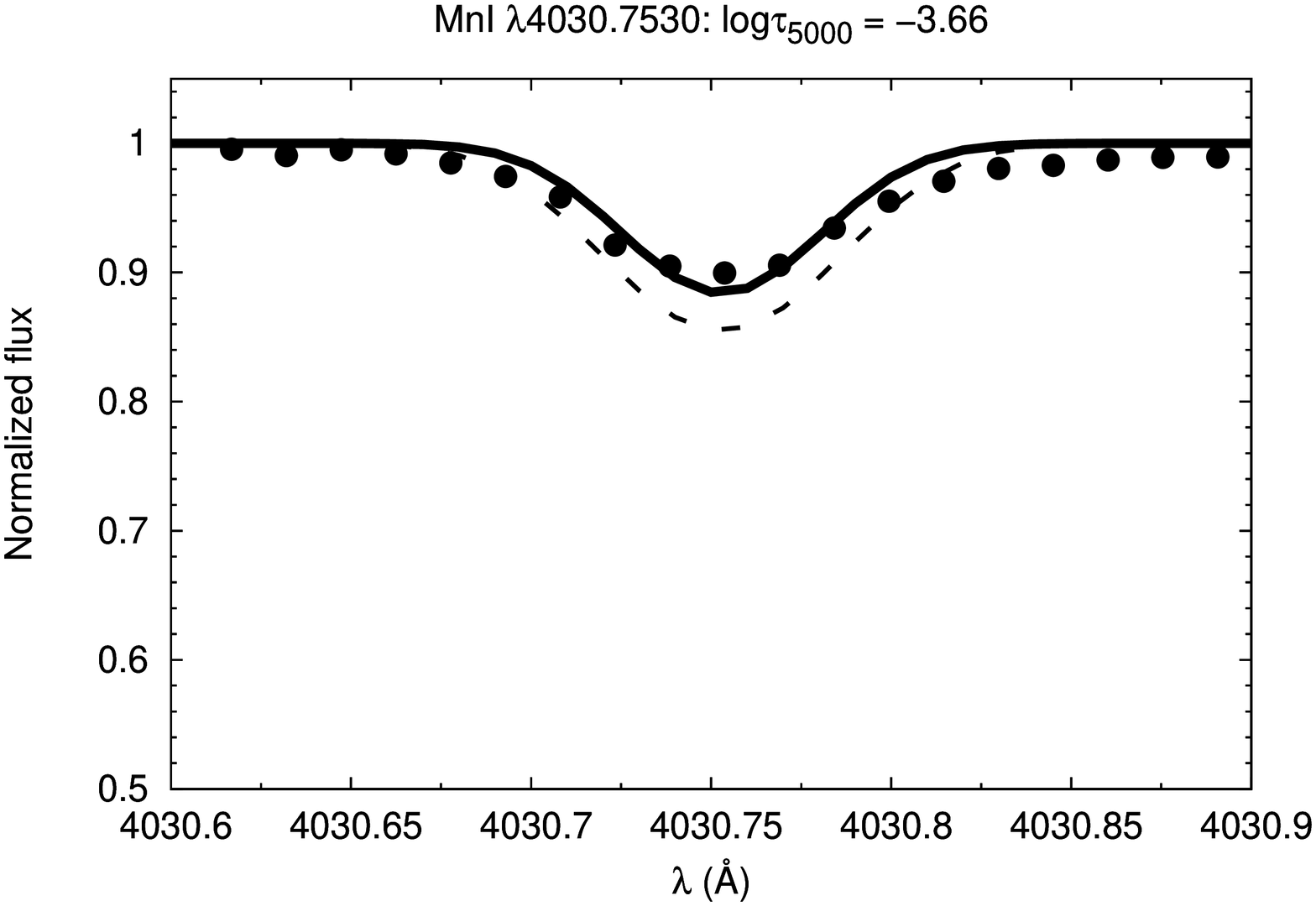}
\includegraphics[scale=0.22,angle=-90]{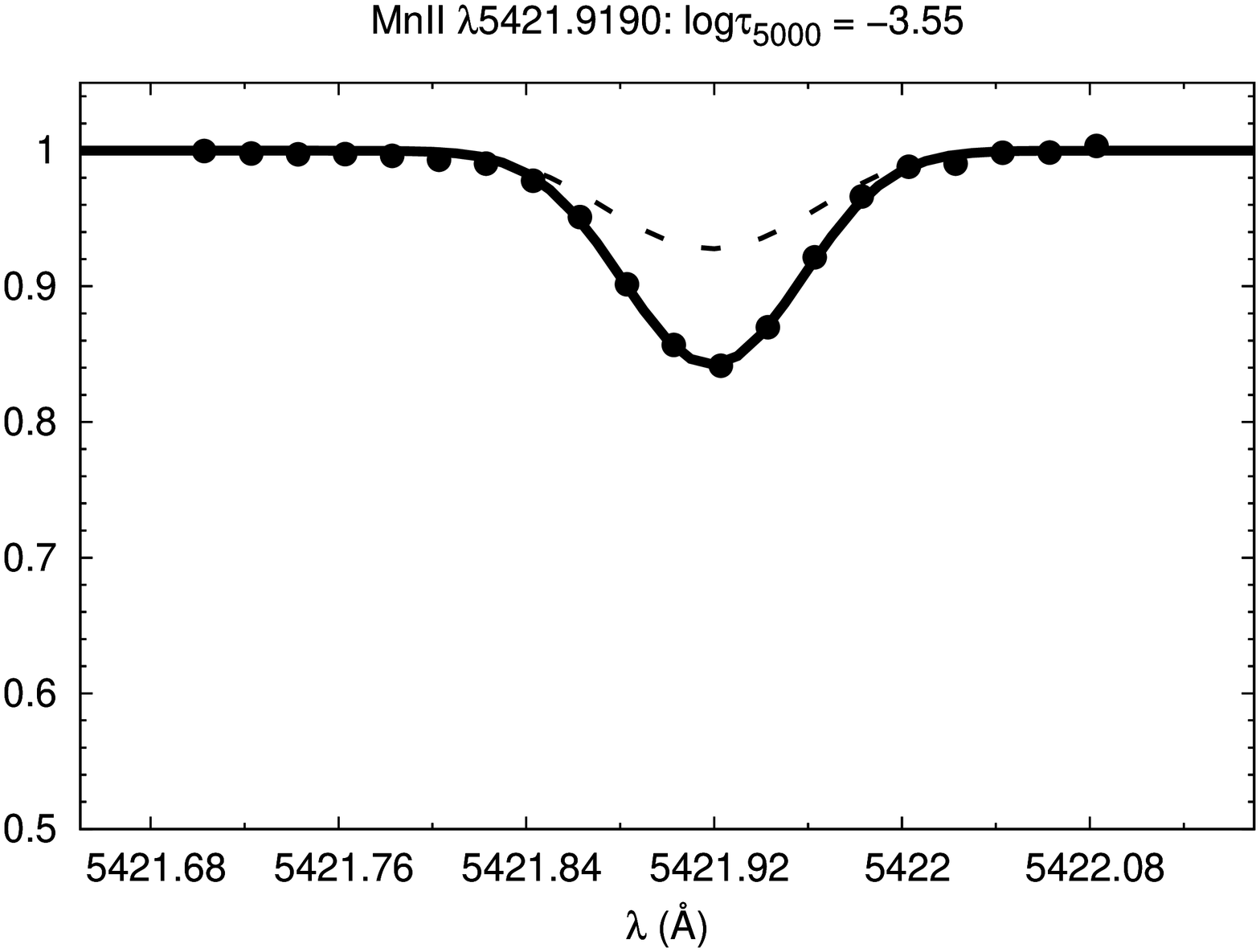}
\includegraphics[scale=0.22,angle=-90]{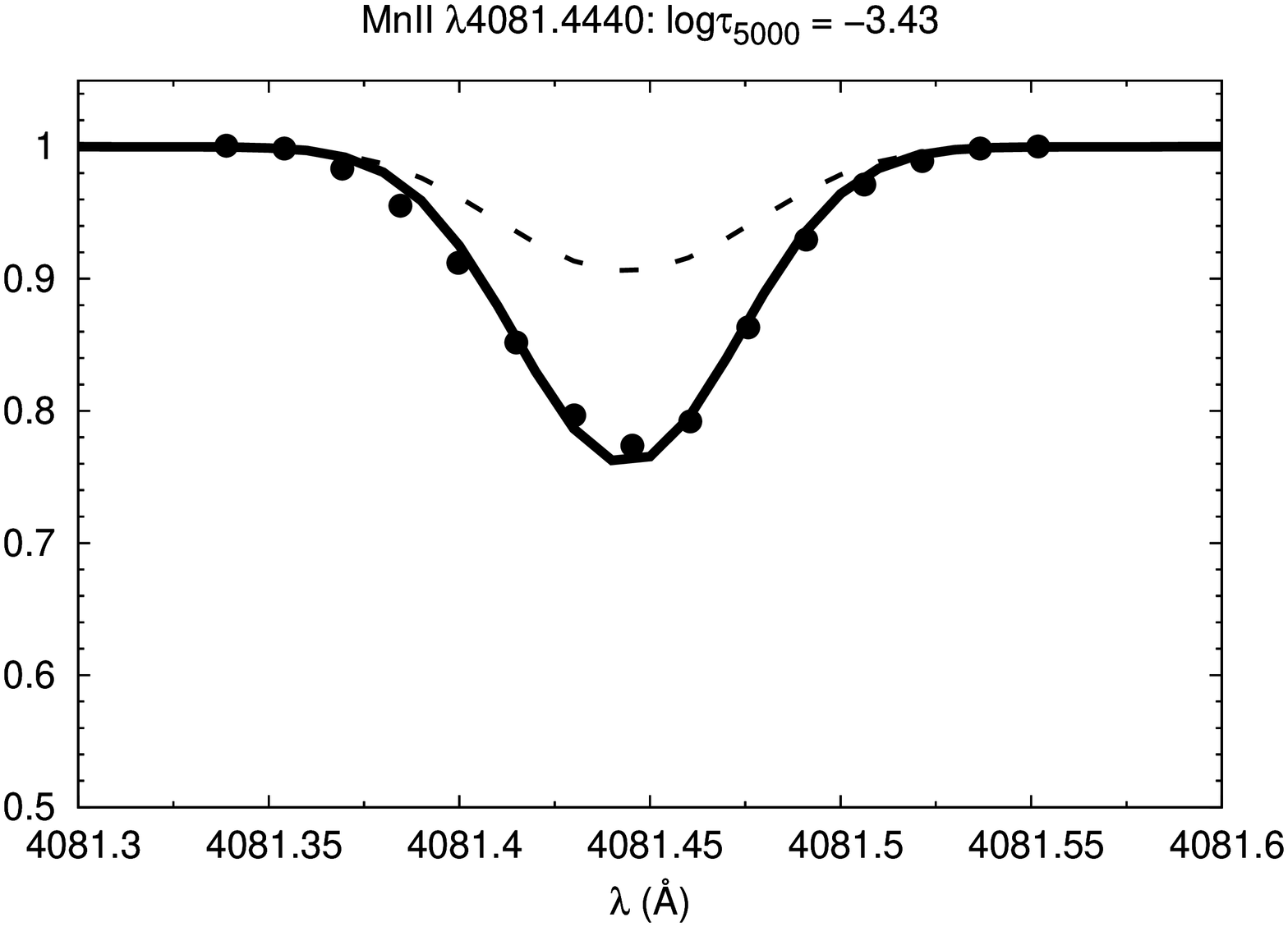}\\
\includegraphics[scale=0.22,angle=-90]{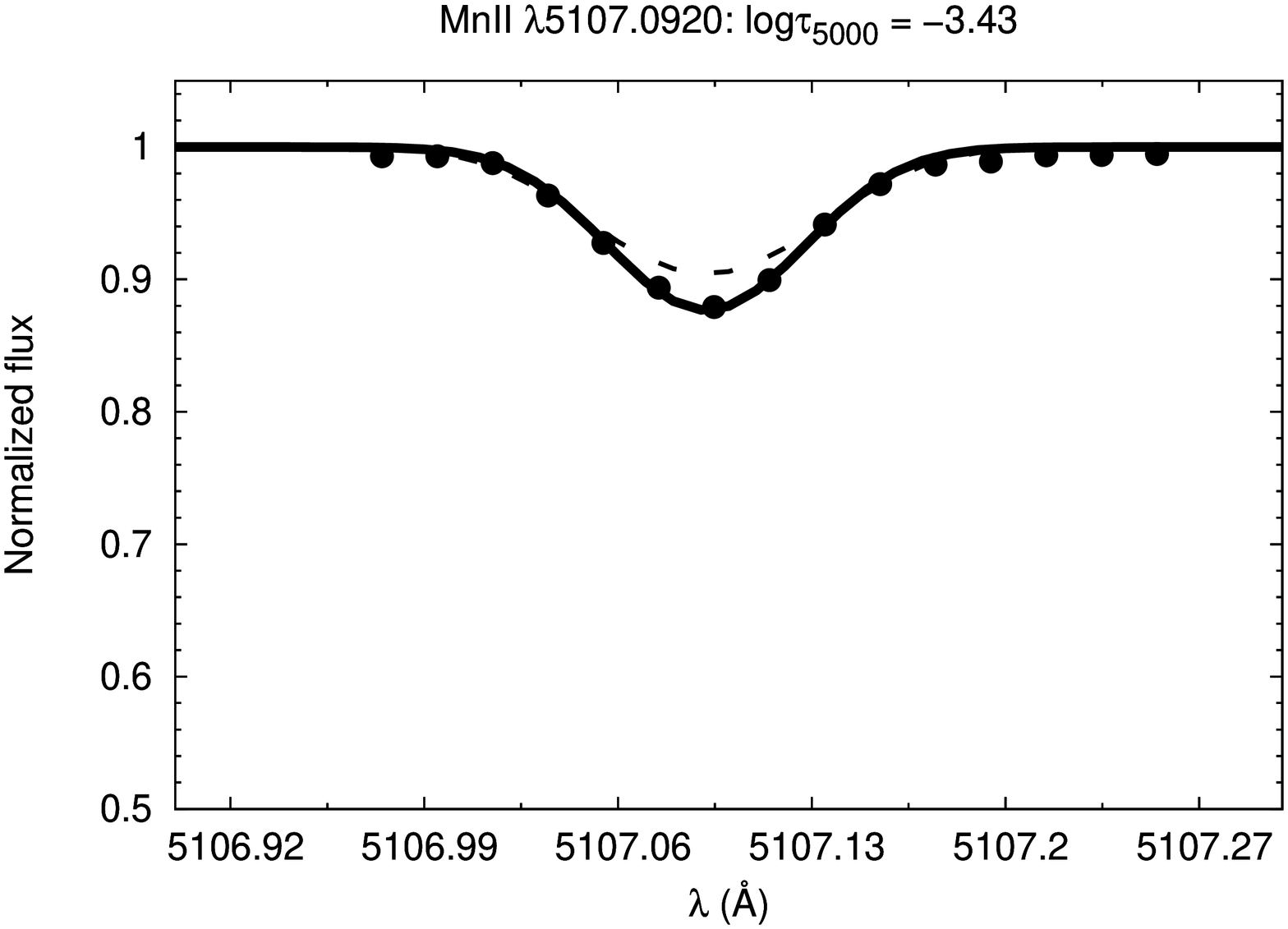}
\includegraphics[scale=0.22,angle=-90]{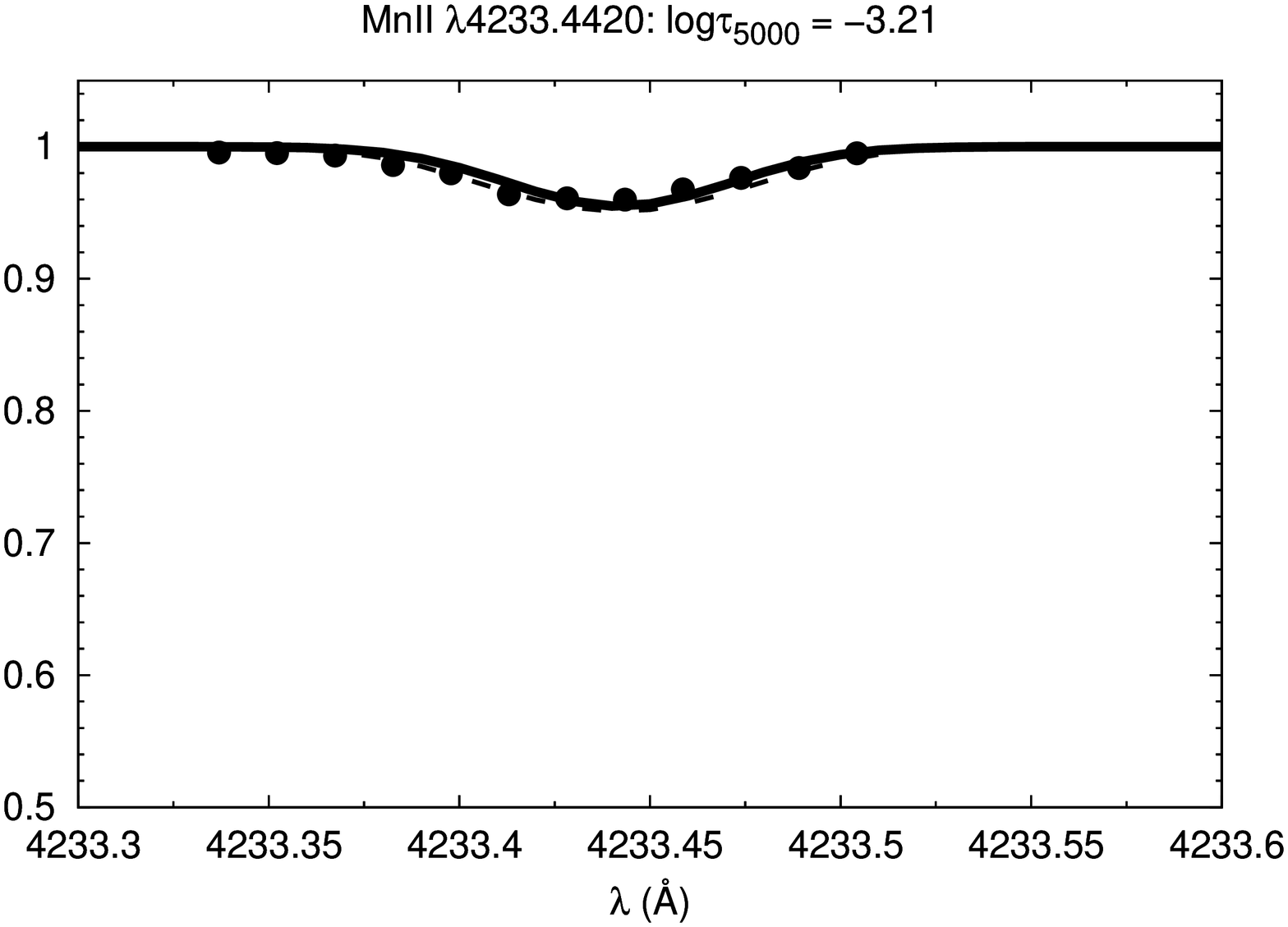}
\includegraphics[scale=0.22,angle=-90]{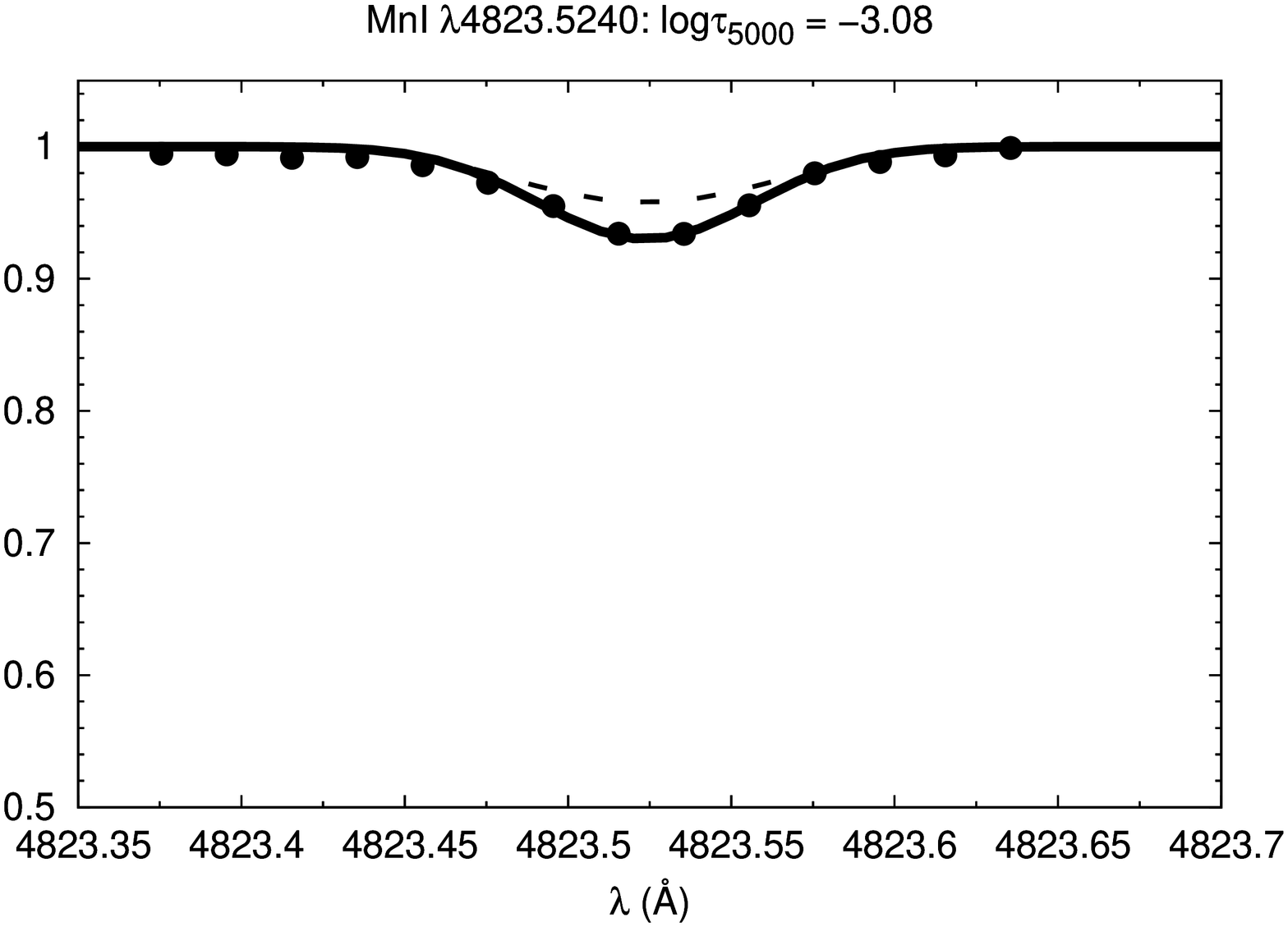}\\
\includegraphics[scale=0.22,angle=-90]{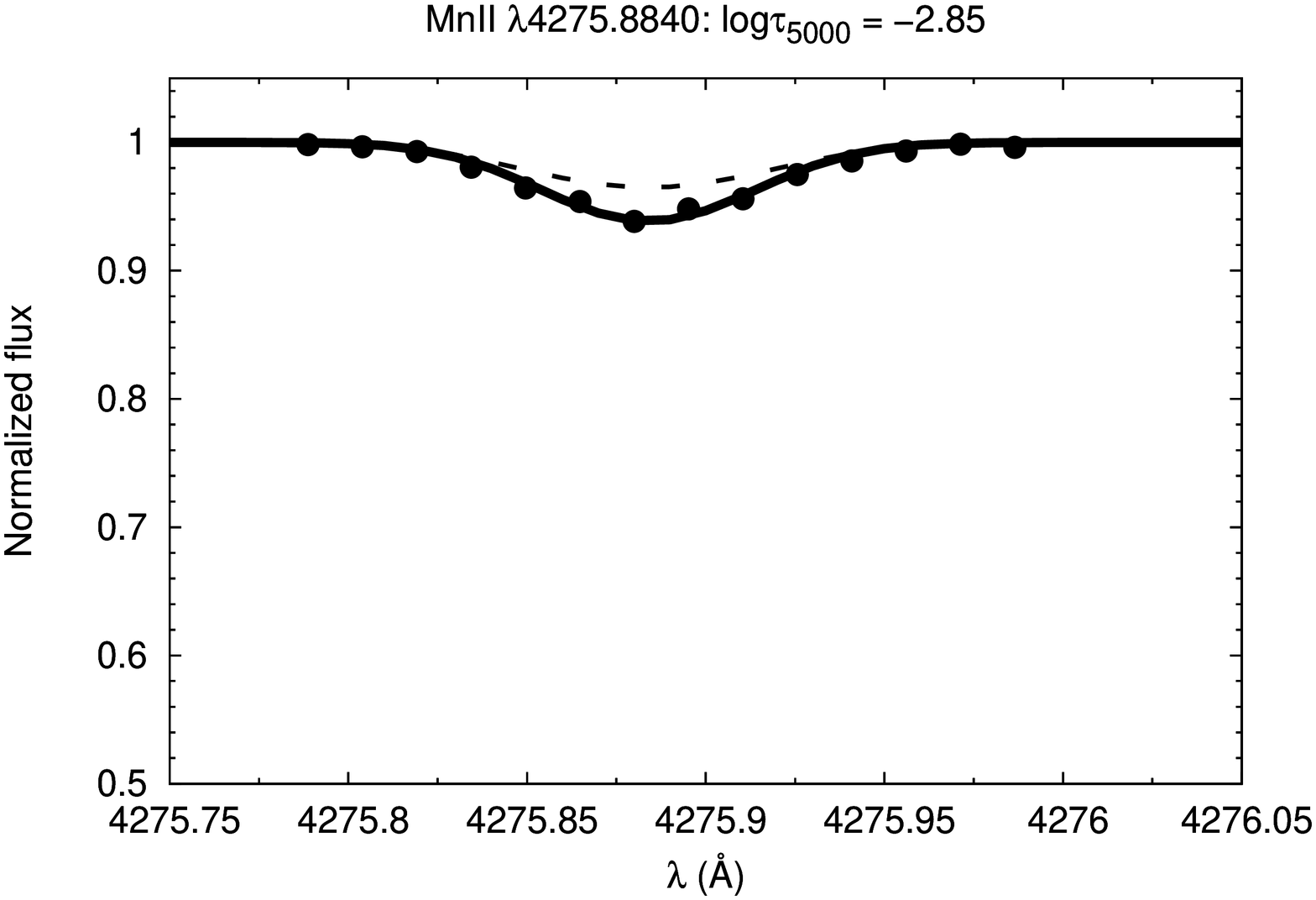}
\includegraphics[scale=0.22,angle=-90]{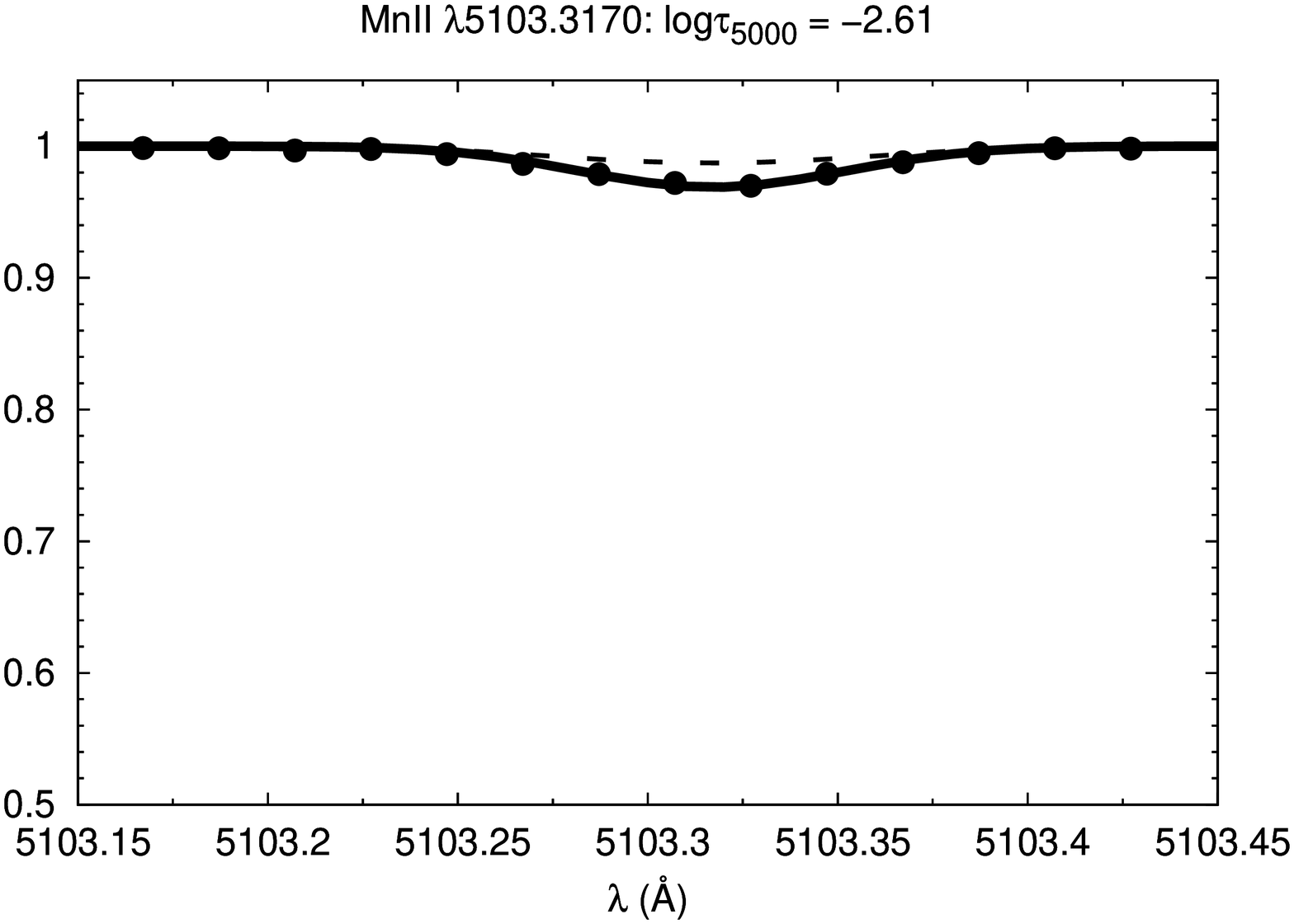}
\includegraphics[scale=0.22,angle=-90]{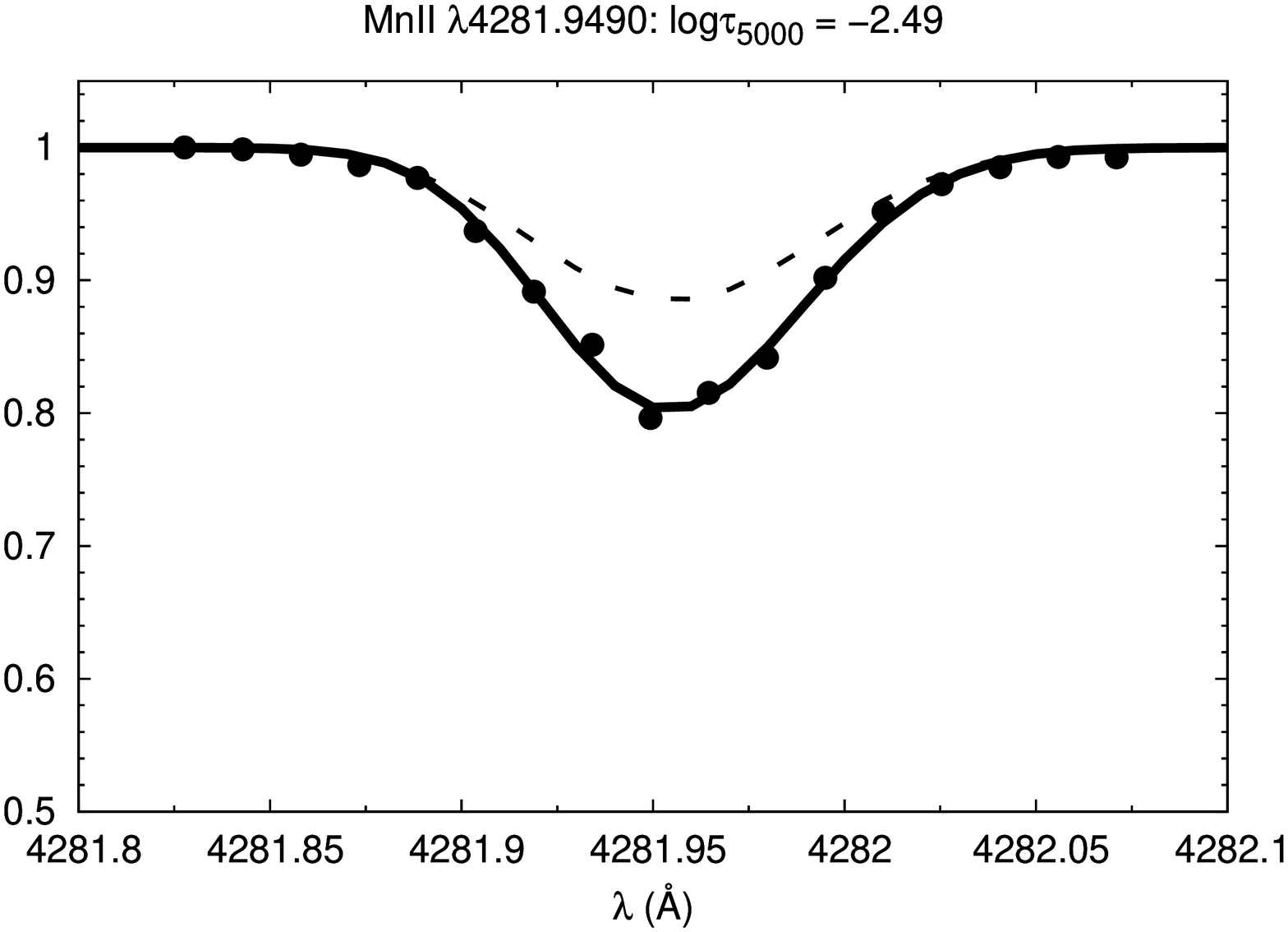}
\caption{A sample of manganese lines observed (dots) for
HD~178065. These lines are fitted by assuming an uniform abundance
(dashed curve) and a stratified one (solid curve). The optical depth
of line core formation is also given for each line.}
\label{fi:raies}
\end{figure*}

To confirm the presence of Mn stratification, several tests have been
undertaken. We began by verifying the effect on the derived
stratification of modifying $T_{\rm eff}$ and the metallicity.

First, we determined the abundance and stratification of Mn using
ATLAS9 model atmospheres with different effective temperatures
centered on the adopted effective temperature of HD~178065 (11~000~K,
11~500~K, 12~500~K and 13~000~K).  Neutral Mn lines are sensitive to
the change of $T_{\rm eff}$. A difference of $\pm 1000$~K (3-4 times
larger than our estimate of the uncertainty associated with $T_{\rm
eff}$) changes the abundance by approximatively $\pm 0.6$~dex. In
contrast, the difference is less than $\sim$0.2~dex for
Mn~{\scriptsize{II}} lines. At each adopted effective temperature, the
slopes remain statistically significant for Mn~{\scriptsize{I}}
(larger than $3 \sigma_{a}$) and Mn~{\scriptsize{II}} (larger than $4
\sigma_{a}$), as well as for all Mn lines (at least $4
\sigma_{a}$). The increase of the Mn abundance is also large enough
(more than 0.5 dex) to be considered numerically significant.

Secondly, we analyzed Mn lines using ATLAS9 atmosphere models with
$T_{\rm eff} = $ 12~000~K, calculated with different metallicities
from $\pm 0.1$ to $\pm 1$~dex. These metallicities correspond to both
deficiencies (negative) and enrichments (positive) of metals relative
to the sun denoted by $[+0.0]$.  The Mn abundances determined with
different metallicities are not very different in comparison to those
obtained with the solar metallicity model. The slopes of the
regression of Mn abundance versus optical depth increase when
metallicity increases.  As an example, for $[-1.0]$, the slope
obtained is $+0.240\pm0.066$, $+0.289\pm0.060$ for $[+0.0]$, and
$+0.326\pm0.058$ for $[+1.0]$.  These remain both statistically and
quantitatively significant for all investigated metallicities.

Thirdly, we have analyzed the Mn lines using a model atmosphere
computed with a different code: an LTE model obtained with the Phoenix
code (Hauschildt et al. 1997). This model is calculated by assuming
$T_{\rm eff}= 12~193$~K, log~$g = 3.54$ while using the abundances
given in Table~\ref{tab:average}.  Mean abundances derived from
Mn~{\scriptsize{I}} and Mn~{\scriptsize{II}} lines are respectively
$-5.17\pm0.18$ and $-5.08\pm0.23$.  No disagreement is observed
between abundances derived from Mn~{\scriptsize{II}} lines using the
two different atmosphere models, while for Mn~{\scriptsize{I}}, the
difference is within the error bars.  The slope $a = +0.296\pm0.061$
obtained from the regression of Mn versus the optical depth using the
Phoenix atmosphere model is close to those obtained with ATLAS9.

\begin{table*}
\begin{minipage}{130mm}
\caption{Results obtained from the simultaneous analysis of manganese
lines of HD~178065 by assuming a two-step model. The variables
$\tau_{1}$ and $\tau_{2}$ are the optical depths where the abundance
of Mn decreases linearly by $\Delta\epsilon$. The quality of the fit
for the uniform and stratified models are respectively represented by
$\chi^{2}_{UM}$ and $\chi^{2}_{SM}$.}
\label{tab:mnfit}
\begin{tabular}{@{} c c c c c c c c c c c @{}}\hline\hline
log ($N_{\rm Mn}$ / $N_{\rm tot}$)$_{1}$&$\Delta\epsilon$&log
                               $\tau_{1}$&log $\tau_{2}$& $V_{\rm r}$&
                               $v~{\rm
                               sin}~i$&$\chi^{2}_{UM}$&$\chi^{2}_{SM}$\\
                               &                   &                 &
                               & (km s$^{-1}$)      & (km s$^{-1}$)
                               &  & \\\hline $-4.47 \pm 0.10$     &
                               $-0.74 \pm 0.21$  & $-2.92 \pm 0.10$ &
                               $-3.79 \pm 0.20$& $34.70 \pm 0.10$ &
                               $1.52 \pm 0.03$  & 13.53 & 4.82\\\hline
\end{tabular}
\end{minipage}
\end{table*}

We therefore conclude that the Mn stratification remains statistically
significant for both ATLAS9 and Phoenix model atmospheres, at all
investigated effective temperatures and metallicities. As the
variation of Mn abundance is substantial, we can confidently claim
that Mn is stratified in the atmospheric layers of HD 178065.

Following the detection and modeling of stratification in magnetic Ap
stars by Babel (1994), vertical abundance profiles were subsequently
parametrized assuming a simple two-step (or two-zone) model (Wade et
al. 2001b; Ryabchikova et al. 2003). In an attempt to characterize the
vertical distribution of Mn in the atmosphere of HD 178065, we have
selected the best line profiles candidates to fit simultaneously using
synthetic profiles computed assuming the two-step model. We assume an
abundance log~$(N_{ion}/N_{tot})_{1}$ deep in the atmosphere, below
optical depth $\tau_{1}$. The abundance in the upper atmospheric
layers (above optical depth $\tau_{2}$) is given by
$\log(N_{ion}/N_{tot})_{1}+\Delta\epsilon$, where $\Delta\epsilon$ is
the difference in abundance between the deeper and upper zones. The
atmospheric layers between $\tau_{1}$ and  $\tau_{2}$ (the "transition
zone") are assumed to have abundances given by a linear interpolation
between that of the deep layers and of the upper layers. In our
fitting of the line profiles, this method operates with six free
parameters (log~$(N_{ion}/N_{tot})_{1}$, $\Delta\epsilon$, $\tau_{1}$,
$\tau_{2}$, $V_{r}$ and $v~\sin~i$) which are simultaneously fitted to
all selected lines.  More details concerning this method are provided
by Khalack et al. (2007).

\begin{figure}
\center
\includegraphics[scale=0.33,angle=-90]{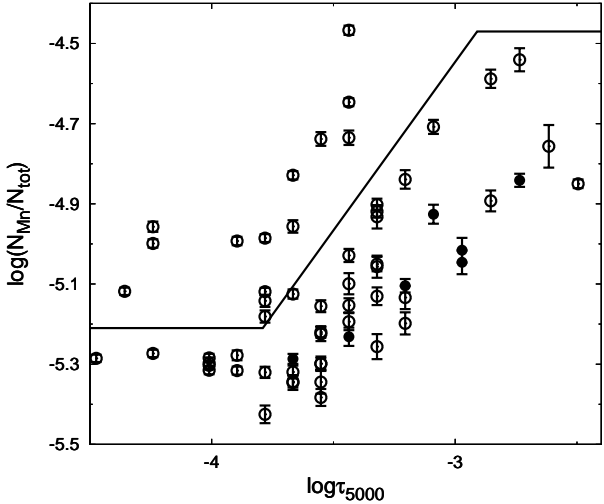}
\vspace{15pt}
\caption{The manganese stratification profile in the atmosphere of
HD~178065. }
\label{fig:tau_strathd178065}
\end{figure}

\begin{figure}
\center
\vspace{270pt}
\includegraphics[scale=0.33,angle=-90]{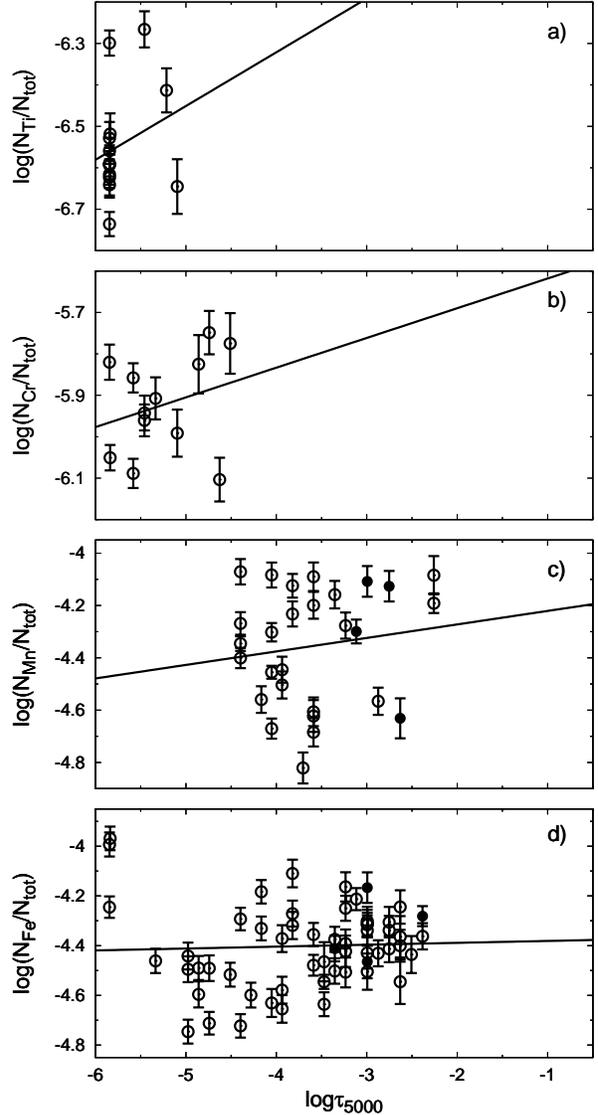}
\vspace{-20pt}
\caption{Same description as in Fig.~\ref{fig:tau_hd71066} but for a) Ti, b) Cr, c)
Mn and d) Fe for HD~221507.}
\label{fig:tau_hd221507}
\end{figure}

\begin{figure}
\center
\vspace{473pt} \includegraphics[scale=0.3,angle=-90]{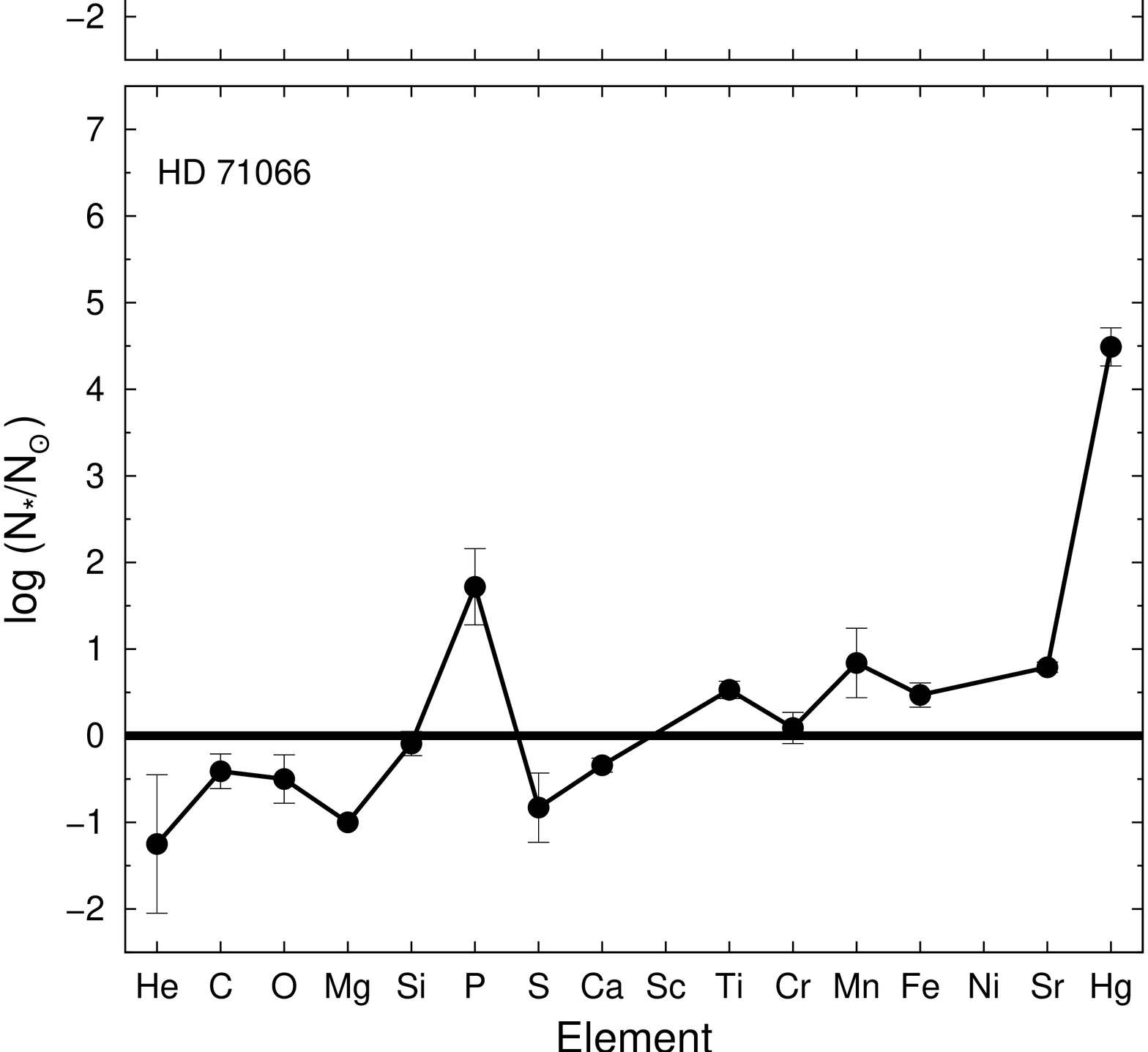}
\vspace{20pt}
\caption{Catalogue of the abundances measured in the atmospheres of
HD~71066, HD~175640, HD~178065 and HD~221507.  These abundances are
relative to the sun, which is represented by the horizontal line.}
\label{fig:abundtotale}
\end{figure}

We simultaneously analyzed the selected Mn line profiles, and the
results of the line synthesis are presented in Table~\ref{tab:mnfit}
and the derived stratification profile is represented in
Fig.~\ref{fig:tau_strathd178065}. Fig.~\ref{fi:raies} shows some
Mn~{\scriptsize{II}} lines formed at different optical depths. These
line profiles are fitted by assuming an uniform abundance (dashed
curve) and a stratified abundance (solid curve). The reduced
$\chi^{2}$ for the uniform and stratified models are summarized
respectively in columns 7 and 8 of Table~\ref{tab:mnfit}. As
represented in Fig.~\ref{fig:tau_strathd178065}, the abundance of Mn
increases deep in the atmosphere of HD~178065 by $0.74\pm0.21$~dex in
the optical depth range between $-3.79\pm0.20$ and $-2.92\pm0.10$. The
abundance of Mn is larger in the lower atmosphere as compared to the
shallower atmosphere. The Mn transition zone observed in HD~178065
dominates the stratified profile. This profile differs strongly from
those of Ap stars where the transition zone has a very small extent
(Ryabchikova et al. 2005). The radial $V_{\rm r}$ and rotational
$v~{\rm sin}~i$ velocities are similar to those adopted from the
individual line analysis discussed in Sect.~3.

\subsection{HD~221507}

In the spectrum of HD~221507, lines of Ti, Cr, Mn and Fe were studied for vertical stratification. 
The linear fit of the abundance derived from 14 Ti lines reveals a large slope. However, this slope is heavily weighted by a few lines which are concentrated close to the log~$\tau_{5000} \approx -5.9$. Also, the lines studied here are formed in a small range of optical depth from log~$\tau_{5000} \approx -5.9$ to $-5$ (see Fig.~$\ref{fig:tau_hd221507}a$). Therefore, it is not possible to conclude from these results that Ti is stratified.

From log~$\tau_{5000} \approx -5.9$ to $-4.5$, the Cr abundance does not show dependence on depth (see Fig.~$\ref{fig:tau_hd221507}b$). The slope obtained ($a~\approx~0.9~\sigma_{a}$) is not statistically significant. Chromium does not show any stratification in the studied optical depth range.

From Mn~{\scriptsize{I}} and Mn~{\scriptsize{II}} lines, we obtain no significant slope from the regression. Manganese does not appear to be stratified in the optical depths between log~$\tau_{5000} \approx -4.4$ and $-2.2$ in HD~221507 (see Fig.~$\ref{fig:tau_hd221507}c$).

The investigated Fe lines are formed between log~$\tau_{5000} \approx
-5.9$ and $-2.3$ (see Fig.~$\ref{fig:tau_hd221507}d$).  The linear fit
of iron abundance reveals a weak slope (see Table~\ref{tab:LSM}) that is not statistically significant.

\begin{figure}
\includegraphics[scale=0.33,angle=-90]{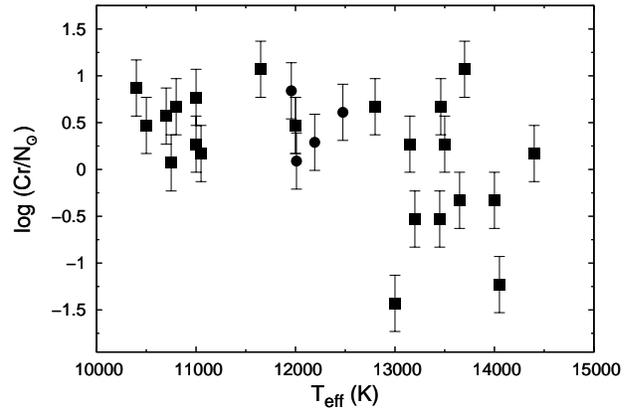}
\caption{The abundance of Cr against the effective temperature of
several HgMn stars. The filled circles and the filled squares
correspond respectively to the abundances derived in our study and
those of Smith \& Dworetsky (1993). }
\label{fig:CrT}
\end{figure}

\section{Discussion and conclusion}

In this paper, we determined the abundances of 16 elements in the
atmospheres of the HgMn stars HD~71066, HD~175640, HD~178065 and
HD~221507. Mean abundances of elements, relative to the sun, are
presented for each star in Fig.~\ref{fig:abundtotale}. The abundances
of most of these elements had never been determined in the past for
these stars. However, for the elements previously investigated and
discussed in Sec.~4, the mean abundances obtained are generally in
good agreement with past studies.

As expected (see Fig.~\ref{fig:abundtotale}), He is underabundant
while Mn and Hg are overabundant in all the investigated
stars. Phosphorus, which is generally overabundant in HgMn stars, is
solar in HD~175640. Titanium and chromium, which are typically present
in their solar abundances in HgMn stars, are strongly enhanced in the
atmosphere of HD~175640. We also notice that the more rapidly rotating
star HD~221507 presents the highest abundances of Sr and Hg of our
sample.

We have examined the dependence of the mean abundances of each element
as a function of the physical parameters $T_{\rm eff}$ and
$v$~sin~$i$. As the results we obtained are derived from only four
data points with highly clustered $T_{\rm eff}$ and $v$~sin~$i$, we
decided to use data from published papers. We used Ryabchikova (1998)
for P and Sr, Smith \& Dworetsky (1993) for Cr, and Smith (1997) for
Hg. We can definitely confirm the correlation of the abundance of Cr
versus $T_{\rm eff}$ (see Fig.~\ref{fig:CrT}), with a large dispersion
of abundances for $T_{\rm eff}\ge13~000$~K (Smith \& Dworetsky,
1993). However, we can not confidently affirm any trends of the
abundance of P with respect to $T_{\rm eff}$, and Sr and Hg with
respect to $v$~sin~$i$.

The synthesis of elements represented by a large number of lines in
our spectra allowed us to track vertical stratification of elements in
the atmospheres of these stars. As shown in Sec.~5, an attempt to
detect stratification for Ti, Cr and Fe in HD~71066, S, Ti, Cr, Mn and
Fe in HD~175640, Ti, Cr, Mn and Fe in HD~178065 and Ti, Cr, Mn and Fe in
HD~221507 was undertaken. Most of these elements' abundances do not
show any significant tendencies in the diagnosed optical depths.

The study of Savanov \& Hubrig (2003) that reports the vertical
stratification of Cr in 10 HgMn stars, is performed in the spectral
region $4812.34-4864.33$~\AA.~They did not evaluate the depth of  line
formation. The spectra used in our study do not appear to be of
sufficient quality to confidently detect the chromium abundance
variation claimed by those authors. Therefore, it is difficult to
compare their data with the results presented here.

An interesting finding is that Fe increases slightly ($\sim~0.25$~dex) with
optical depth for the four stars (HD~71066, HD~175640, HD~178065 and HD~221507). This systematic trend for Fe
might suggest stratification. However the increase in abundance is too
small to confidently conclude to the detection of stratification.

Strong evidence of Mn stratification in the atmosphere of HD~178065 is
observed. The abundance of Mn increases by $\sim~0.7$~dex in diagnosed
optical depth range. The simultaneous fit of lines assuming a two-step
model also shows that Mn increases towards the deep atmosphere.  If
such stratification can be confirmed (for Mn or for other elements),
this would provide important new quantitative constraints on diffusion
in the atmosphere of HgMn stars. Such observed vertical stratification
profiles could then be compared to theoretical model atmospheres which
include elemental stratification such as those of Hui-Bon-Hoa et
al. (2000).

Differences observed between the abundance of neutral and ionized Mg
ions are greater than $0.1$~dex in HD~178065 and
HD~221507. The largest disagreements are observed between Hg~{\scriptsize{I}} and
Hg~{\scriptsize{II}} in HD~175640, HD~178065 and HD~221507, where they
are greater than $0.25$~dex. This result remains the same even when not
using the isotopic data for the Hg~{\scriptsize{II}}~$\lambda$~3984
line shown in Fig.~\ref{fit:HgISS}.  The fact that differences are
observed for some stars and not for others lets us to conclude that
they may not be only attributed to uncertainties surrounding atomic
data and/or small number of lines observed. The sources of these
differences might be attributed to NLTE effects. Takada-Hidai (1991)
showed that neutral ions are more affected by NLTE corrections than
ionized ones in HgMn stars. However, NLTE corrections are smaller for
the heavier ions (Takada-Hidai 1991). Basing on the aforementioned
paper, the magnitudes of NLTE corrections for some elements are
$-0.4$~dex for Be~{\scriptsize{II}}, from $-2.00$~dex to $0.00$~dex
for O~{\scriptsize{I}}, $-0.02$~dex for Mg~{\scriptsize{I}},
$-0.04$~dex for Mg~{\scriptsize{II}}, from $-0.45$~dex to $+0.70$~dex
for Ca~{\scriptsize{II}}, $+0.20$~dex for Sr~{\scriptsize{II}} and
from $-0.3$~dex to $+0.3$~dex for Ba~{\scriptsize{II}}. The
corrections are given to indicate possible uncertainties in our
results. After the discovery of Ga ionization anomalies in HgMn stars
by Smith (1995), this author suggested stratification as the possible
cause of these anomalies. 

In this paper, we reported for the first time the detection of Mn stratification in a HgMn type star. However, the attempt to detect element stratification failed for most of the investigated elements.
The investigation of HgMn stars could be complicated by several physical process, which should be taken into account.
Sigut (2001) suggested that the emission lines present in spectra of some HgMn stars are photospheric in nature.
After the first discovery of Mn~{\scriptsize{II}} emission lines in the spectra of the HgMn star HD~186122 (46 Aql) by Sigut et al. (2000), emission lines of several elements were also discovered (Wahlghren \& Hubrig 2000). These authors observed titanium and chromium emission lines in the spectra of HD~71066, HD~175640 and HD~178065. As interpreted by Sigut (2001), the presence of emission lines is caused by interlocked NLTE effects in the region where the element is stratified. When doing LTE analysis such as done in this paper, the presence of such emission lines increases the uncertainty of their results.

Possible presence of spots and weather in the atmosphere of HgMn stars also add complications to the investigation of
these stars. The variability of Hg~{\scriptsize{II}}~$\lambda 3984$ in the spectra HgMn stars is interpreted as
inhomogeneous distribution of this element on the stellar surface (Adelman et al. 2002; Kochukhov et al. 2005).
To properly study this line, isotopic consideration is critical.
Several line profile variations are found in other HgMn stars, such as the strong line variations of Pt, Hg, Sr, Y, Zr,
He and Nd in the eclipsing binary AR~Aur by Hubrig et al. (2006).

We notice here the absence of the mercury line (Hg~{\scriptsize{II}} $\lambda5677$) in the spectrum of HD~175640. This means that the conditions required for the formation of the absorption or the emission Hg~{\scriptsize{II}}~$\lambda5677$ line are not met in the specific layers. On the other hand, it could also mean that the absorption line is formed in deeper layers, but the emission line forms in upper layers and fills in the line. This adds a constraint to the hypothesis that NLTE effects probably act in the atmosphere of HD~175640.
In future studies, improvements must be brought to the characterization of the optical depth of line formation and inclusion of
NLTE effects. The study of spectral regions covering the UV and IR may allow to diagnose deeper and
shallower regions of the atmospheres. Also, spectra with a high S/N and accurate atomic data might permit us to detect weaker stratification.

\section*{Acknowledgments}
We thank ESO for the availability of the spectra used in this work.
We would like to thank J.D. Landstreet and V. Tsymbal for useful discussions, as well as spectra and
codes they have made available to us.
Thanks to R\'eseau Qu\'eb\'ecois de Calcul de Haute Performance (RQCHP) for computationnal ressources. This research was partially supported by NSERC.
GAW is supported by the Academic Research Programme of the Canadian Department of National Defence. FL thanks the Facult\'e des \'Etudes Sup\'erieures et de la Recherche de l'Universit\'e de Moncton for financial support. We thank the referee for the helpful comments.

\bsp
\newpage

\begin{table*}
\begin{minipage}{100mm}
\center
\caption{The hyperfine structure of the Mn~{\scriptsize{II}}~$\lambda$~4206 line taken from Holt et al. (1999).}
\label{tab:MnHFS}
\begin{tabular}{ c c }\hline
\hline
$\lambda$~(\AA) &  log~$gf$ \\\hline
4206.2960            &  -4.5828  \\
4206.2980            &  -3.3653 \\
4206.2994            &  -2.6963\\
4206.3111            &  -4.4367\\
4206.3136            &  -3.1746\\
4206.3156            &  -2.5872\\
4206.3306            &  -4.5238\\
4206.3337            &  -3.1211\\
4206.3362            &  -2.4777\\
4206.3545            &  -4.8517\\
4206.3582            &  -3.1626\\
4206.3613            &  -2.3729\\
4206.3872            &  -3.3534\\
4206.3909            &  -2.2742\\
4206.4249            &  -2.1814\\\hline
\end{tabular}
\end{minipage}
\end{table*}

\begin{table*}
\begin{minipage}{100mm}
\center
\caption{The isotopic composition of the Hg~{\scriptsize{II}}~$\lambda$~3984 line taken from Dolk et al. (2003).}
\label{tab:HgIS}
\begin{tabular}{ c c c }\hline
\hline
Isotope      & $\lambda$~(\AA) &  log~$gf$ \\\hline
$^{196}$Hg & 3983.7710            & -4.550 \\
$^{198}$Hg & 3983.8386     &     -2.730\\
$^{199a}$Hg & 3983.8437     &   -2.880  \\
$^{199b}$Hg & 3983.8527     &    -2.740 \\
$^{200}$Hg & 3983.9124     &     -2.370\\
$^{201a}$Hg & 3983.9316     &    -3.050 \\
$^{201b}$Hg & 3983.9414     &    -2.810 \\
$^{202}$Hg & 3983.9932     &     -2.250\\
$^{204}$Hg & 3984.0724     &     -2.880\\\hline
\end{tabular}
\end{minipage}
\end{table*}

\begin{table*}
\begin{minipage}{80mm}
\center
\caption{Isotopic mixture of the Hg~{\scriptsize{II}}~$\lambda$~3984 line for the studied stars.}
\label{tab:HgISC}
\begin{tabular}{l l l l l r c r}\hline\hline
HD                   &  \multicolumn{7}{c}{Isotopic mixture}\\\hline
                         &   196      &   198  & 199 & 200 & 201 & 202 & 204 \\\hline
71066             &    1.1       &    4.0  &  3.4 & 15.2 & 66.9 & 8.1 & 1.3 \\
175640           &    0.0     &     9.6    & 8.4 & 29.2 &   6.2 &  43.3 & 3.3 \\
178065           &   0.0      &    5.7     &  15.2 & 29.9 &  2.0 & 24.0& 23.2 \\
221507           &  0.0    &  10.0         &  8.5  &   10.0 &  6.6 &  30.2 &  34.7\\\hline
\end{tabular}
\end{minipage}
\end{table*}

\newpage

\begin{table*}
\caption{Atomic data used for the spectral synthesis. The columns denoted by 1, 2, 3 and 4 are respectively the abundances measured in the atmospheres of HD~71066, HD~175640, HD~178065 and HD~221507. The source of the atomic data is mostly VALD, unless some other sources are specified in column 10. The damping parameters log~$\gamma_{\rm rad}$ and log~$\gamma_{\rm St}$ which are not available are represented by a horizontal line.}
\label{tab:donnees_atomiques}
\begin{center}
\begin{tabular}{@{}l r c c c c c c c c @{}}\hline\hline
$\lambda$ (\AA) & $E_{\rm i}$ (eV) & log $gf$ & log $\gamma_{\rm rad}$ & log $\gamma_{\rm St}$ & 1 & 2 & 3 & 4 & Sources \\\hline

& & & & C~{\scriptsize{II}}& & & \\
3918.9702 & 16.3320 & -0.533 & 8.980  &  -4.930 & -3.96    & -3.86 & -3.78  &           &  \\
4267.2590 &  18.0460 &0.717  & 9.390  & -4.760  & -3.82     &          &            &  -3.92 &   \\\hline

& & & & O~{\scriptsize{I}}& & & \\
3947.2928 & 9.1460  & -2.096 &  6.660 & -4.700  & -3.74 &       & -3.54  &       &\\
3947.4827 & 9.1460  & -2.244 &  6.660 & -4.700  & ...   &       & ...    &       & \\
3947.5827 & 9.1460  & -2.467 &  6.660 & -4.700  & ...   &       & ...    &       & \\
\hline
\end{tabular}
\end{center}
\end{table*}

\begin{table*}{{\bf Table \ref{tab:donnees_atomiques} - }continued.}
\begin{center}
\begin{tabular}{@{}l r c c c c c c c c @{}}\hline\hline
$\lambda$ (\AA) & $E_{\rm i}$ (eV) & log $gf$ & log $\gamma_{\rm rad}$ & log $\gamma_{\rm St}$ & 1 & 2 & 3 & 4 & Sources \\\hline

& & & & P {\scriptsize{II}}& & & \\
5296.0770 & 10.8021 & -0.16 & --- & --- & -4.86 &        & -5.09 &       & NIST \\
5344.7290 & 10.7367 & -0.39 & --- & --- &       & -6.30  & -5.14 & -5.60 & NIST \\
5409.7330 & 10.7549 & -0.39 & --- & --- & -4.88 &        & -5.16 & -5.36 & NIST \\
5425.8800 & 10.8021 &  0.18 & --- & --- & -4.85 & -6.29  & -5.06 &       & NIST \\
6034.0400 & 10.7367 & -0.22 & --- & --- &       &        & -5.05 & -4.77 & NIST \\
6087.8400 & 10.7549 & -0.35 & --- & --- &       &        & -5.25 &       & NIST \\
\hline

\end{tabular}
\end{center}
\end{table*}

\begin{table*}{{\bf Table \ref{tab:donnees_atomiques} - }continued.}
\begin{center}
\begin{tabular}{@{}l r c c c c c c c c @{}}\hline\hline
$\lambda$ (\AA) & $E_{\rm i}$ (eV) & log $gf$ & log $\gamma_{\rm rad}$ & log $\gamma_{\rm St}$ & 1 & 2 & 3 & 4 & Sources \\\hline

& & & & Ti {\scriptsize{II}}& & & \\
4418.3300 & 1.2370 & -1.990 &   8.471  &   -6.601 &       & -5.90  &       &       & \\
4421.9380 & 2.0610 & -1.580 &   8.412  &   -6.587 &       & -5.97  &       &       & \\
4441.7340 & 1.1800 & -2.270 &   8.225  &   -6.694 &       & -5.95  &       &       & \\
4443.7940 & 1.0800 & -0.700 &   8.199  &   -6.509 & -6.52 & -5.76  & -6.59 & -6.59 & \\
4450.4820 & 1.0840 & -1.510 &   8.199  &   -6.502 & -6.58 & -5.86  & -6.60 & -6.52 & \\
4464.4500 & 1.1610 & -1.810 &   8.217  &   -6.713 & -6.57 & -5.86  & -6.53 &       & \\
\hline

\end{tabular}
\end{center}
\end{table*}

\begin{table*}{{\bf Table \ref{tab:donnees_atomiques} - }continued.}
\begin{center}
\begin{tabular}{@{}l r c c c c c c c c @{}}\hline\hline
$\lambda$ (\AA) & $E_{\rm i}$ (eV) & log $gf$ & log $\gamma_{\rm rad}$ & log $\gamma_{\rm St}$ & 1 & 2 & 3 & 4 & Sources \\\hline

& & & & Cr {\scriptsize{II}}& & & \\
4558.6500 & 4.0730 & -0.449 & 8.410  &   -6.638 & -6.19 & -5.50  &       &      & \\
4565.7400 & 4.0420 & -1.820 & 8.410  &   -6.720 &       & -5.56  & -6.07 &       & \\
4588.1990 & 4.0710 & -0.627 & 8.410  &   -6.656 &       & -5.43  &       & & \\
4592.0490 & 4.0740 & -1.221 & 8.410  &   -6.656 & -6.35 & -5.59  & -6.13 & -6.09 & \\
4616.6290 & 4.0720 & -1.361 & 8.410  &   -6.653 & -6.31 & -5.55  & -6.09 & -5.94 & \\
4618.8030 & 4.0740 & -0.840 & 8.410  &   -6.653 & -6.30 & -5.48  & -6.06 & -5.82      & \\
\hline

\end{tabular}
\end{center}
\end{table*}

\begin{table*}{{\bf Table \ref{tab:donnees_atomiques} - }continued.}
\begin{center}
\begin{tabular}{@{}l r c c c c c c c c @{}}\hline\hline
$\lambda$ (\AA) & $E_{\rm i}$ (eV) & log $gf$ & log $\gamma_{\rm rad}$ & log $\gamma_{\rm St}$ & 1 & 2 & 3 & 4 & Sources \\\hline

& & & & Mn {\scriptsize{II}}& & & \\
3941.2310 & 5.4370 & -1.147 & 8.348    & -6.652  &       &        &       &  & \\
4081.4440 & 6.1110 & -2.239 & 8.465    & -6.618  &       &        & -4.65 & & \\
4085.0440 & 5.3730 & -2.543 & 8.420    & -6.680  &       &        & -5.38 &       & \\
4085.3900 & 5.5480 & -2.561 & 8.458    & -6.661  &       &        & -5.30 &       & \\
4085.4050 & 5.3800 & -2.517 & 8.420    & -6.680  &       &        &  ...  &       & \\
4140.4630 & 5.5670 & -2.455 & 8.489    & -6.654  &       & -4.40  & -5.22 & -4.68     & \\
\hline

\end{tabular}
\end{center}
\end{table*}

\begin{table*}{{\bf Table \ref{tab:donnees_atomiques} - }continued.}
\begin{center}
\begin{tabular}{@{}l r c c c c c c c c @{}}\hline\hline
$\lambda$ (\AA) & $E_{\rm i}$ (eV) & log $gf$ & log $\gamma_{\rm rad}$ & log $\gamma_{\rm St}$ & 1 & 2 & 3 & 4 & Sources \\\hline

& & & & Mn {\scriptsize{II}}& & & \\
4717.2640 & 6.5360 & -1.856 & 8.684    & -6.640  &       & -4.55  &       & -4.62      & \\
4727.4610 & 2.9200 & -2.017 & 8.415    & -6.635   &      &             &       & -4.19     & \\
4730.3950 & 5.3730 & -2.147 & 8.417    & -6.634  &       &        & -4.75 & -4.45      & \\
4734.1360 & 5.3730 & -2.879 & 8.415    & -6.635  &       &        &       & -4.28 & \\
4770.3510 & 5.3980 & -2.290 & 8.418    & -6.633  &       &        &       & -4.23 & \\
4784.6250 & 6.5730 & -1.505 & 8.684    & -6.640  &       & -4.51  & -5.28 & -4.44      & \\
\hline

\end{tabular}
\end{center}
\end{table*}

\begin{table*}{{\bf Table \ref{tab:donnees_atomiques} - }continued.}
\begin{center}
\begin{tabular}{@{}l r c c c c c c c c @{}}\hline\hline
$\lambda$ (\AA) & $E_{\rm i}$ (eV) & log $gf$ & log $\gamma_{\rm rad}$ & log $\gamma_{\rm St}$ & 1 & 2 & 3 & 4 & Sources \\\hline

& & & & Fe {\scriptsize{I}}& & & \\
3927.9199             & 0.1100 & -1.522  & 7.152   & -6.324 & -4.04 &        &       &       & \\
3930.2967             & 0.0870 & -1.491  & 7.161   & -6.324 & -4.07 &        &       &       & \\
4005.2420             & 1.5570 & -0.610  & 8.009   & -6.213 & -3.99 &        &       &      & \\
4045.8124             & 1.4850 &  0.280  & 8.009   & -6.205 & -3.95 & -4.93  & -4.67 &  & \\
4071.7380             & 1.6080 & -0.022  & 8.009   & -6.212 & -4.01 & -4.94  & -4.66 &       & \\
4132.0582             & 1.6080 & -0.675  & 8.013   & -6.049 & -3.98 &        &       &       &\\
\hline

\end{tabular}
\end{center}
\end{table*}

\begin{table*}{{\bf Table \ref{tab:donnees_atomiques} - }continued.}
\begin{center}
\begin{tabular}{@{}l r c c c c c c c c @{}}\hline\hline
$\lambda$ (\AA) & $E_{\rm i}$ (eV) & log $gf$ & log $\gamma_{\rm rad}$ & log $\gamma_{\rm St}$ & 1 & 2 & 3 & 4 & Sources \\\hline

& & & & Fe {\scriptsize{II}}& & & \\
4319.4130  & 7.6530  & -2.119 & 9.045    & -5.844 & -3.88 &       &       &       & \\
4319.6800  & 7.8450  & -1.758 & 9.009    & -5.702 & -3.85 &       & -4.48 &       & \\
4321.3090  & 7.8690  & -1.830 & 9.026    & -5.708 & -3.86 &       &       &       & \\
4325.4360  & 6.1380  & -2.337 & 8.617    & -6.580 & -3.96 &       &       &       & \\
4325.5400  & 6.0880  & -2.307 & 8.670    & -6.586 & ...   &       &       &       & \\
4354.3440  & 7.6530  & -1.395 & 9.042    & -5.850 & -4.01 &       &       &       & \\
\hline

\end{tabular}
\end{center}
\end{table*}

\begin{table*}{{\bf Table \ref{tab:donnees_atomiques} - }continued.}
\begin{center}
\begin{tabular}{@{}l r c c c c c c c c @{}}\hline\hline
$\lambda$ (\AA) & $E_{\rm i}$ (eV) & log $gf$ & log $\gamma_{\rm rad}$ & log $\gamma_{\rm St}$ & 1 & 2 & 3 & 4 & Sources \\\hline

& & & & Fe {\scriptsize{II}}& & & \\
4953.9870  & 5.5710  & -2.757 & 8.528    & -6.608 &       &       & -4.57 &       & \\
4984.4880  & 10.3290 &  0.011 & 9.011    & -5.367 &       & -4.81 & -4.56 & -4.34 & \\
4984.5770  & 10.4140 & -1.257 & 9.112    & -5.424 &       & ...   &       &       & \\
4990.5090  &10.3290  &  0.177 & 8.988    & -5.350 & -3.81 &       & -4.57 & -4.32 & \\
4991.4400  &10.2730  & -0.571 & 9.100    & -5.551 &       &       & -4.49 &       & \\
4993.3580  & 2.8070  & -3.640 & 8.484    & -6.588 & -3.91 &       & -4.63 & & \\
\hline

\end{tabular}
\end{center}
\end{table*}

\begin{table*}{{\bf Table \ref{tab:donnees_atomiques} - }continued.}
\begin{center}
\begin{tabular}{@{}l r c c c c c c c c @{}}\hline\hline
$\lambda$ (\AA) & $E_{\rm i}$ (eV) & log $gf$ & log $\gamma_{\rm rad}$ & log $\gamma_{\rm St}$ & 1 & 2 & 3 & 4 & Sources \\\hline

& & & & Fe {\scriptsize{II}}& & & \\
5214.4890  &10.5030  & -0.532 & 9.006    & -5.442 & -3.78 &       &       &       & \\
5215.7630  & 8.1450  & -1.522 & 9.052    & -5.822 & -3.71 &       &       &       & \\
5215.8440  & 10.3810 & -0.227 & 9.221    & -4.959 & ...   &       &       &       & \\
5216.8540  & 10.5030 &  0.390 & 8.956    & -5.367 & -3.80 &       &       &       & \\
5216.8630  & 10.4800 &  0.608 & 8.935    & -5.383 & ...   & -4.74 &       &       & \\
5224.4110  & 10.4140 & -0.572 & 9.088    & -5.411 & -3.80 &       &       &       & \\
\hline

\end{tabular}
\end{center}
\end{table*}

\begin{table*}{{\bf Table \ref{tab:donnees_atomiques} - }continued.}
\begin{center}
\begin{tabular}{@{}l r c c c c c c c c @{}}\hline\hline
$\lambda$ (\AA) & $E_{\rm i}$ (eV) & log $gf$ & log $\gamma_{\rm rad}$ & log $\gamma_{\rm St}$ & 1 & 2 & 3 & 4 & Sources \\\hline

& & & & Fe {\scriptsize{III}}& & & \\
4419.5960 & 8.2410 & -2.218 & 9.061 & -6.858 & -3.89 &   &  &  & \\
5156.1110 & 8.6410 & -2.018 & 9.061 & -6.858 & -3.79 &   &  &  & \\\hline
& & & & Ni {\scriptsize{II}}& & & \\
3849.5540 & 4.0320 & -1.878 & 8.741 & -6.616 &       &        & -6.61 &       & \\
4067.0310 & 4.0290 & -1.835 & 8.653 & -6.617 &       & -6.46  & -6.59 &       & \\
5021.4710 & 12.4370& 0.920  & 8.944 & -5.401 &       &        &       & -6.14 & \\\hline
& & & & Sr {\scriptsize{II}}& & & \\
4077.7090 & --- &  0.167 & --- & -7.710 & -8.35 & -8.61  & -9.21 & -6.00 & \\\hline
& & & & Hg {\scriptsize{I}}& & & \\
4358.3230 &  4.8870 & -0.383 & --- & --- &  -6.58 &         &        &       & \\
5460.7310 &  5.4610 & -0.137 & --- & --- &  -6.19 &  -6.13  & -6.36  & -4.59 & \\\hline
& & & & Hg {\scriptsize{II}}& & & \\
3983.9312 & 4.4030  & -1.730 & --- & --- & -6.76 &  -6.38   & -6.75  & -5.70 & Dolk\\
5677.1050 & 13.0858 & 0.820  & --- & --- & -6.29 &          & -7.02  & -5.51 & NIST\\\hline

\end{tabular}
\end{center}
\end{table*}

\label{lastpage}

\begin{thebibliography}{99}
\bibitem[\protect\citeauthoryear{Abt et al.}{1972}]{b1} Abt H.A., Chaffee F.H., Suffolk G., 1972, ApJ, 175, 779
\bibitem[\protect\citeauthoryear{Adelman et al.}{2002}]{b2} Adelman S.J., Gulliver A.F., Kochukhov O.P., Ryabchikova T.A., 2002, ApJ, 575, 449
\bibitem[\protect\citeauthoryear{Adelman}{1994}]{b3} Adelman S.J., 1994, MNRAS, 266, 97
\bibitem[\protect\citeauthoryear{Alecian}{1982}]{b4} Alecian G., 1982, A\&A, 107, 61
\bibitem[\protect\citeauthoryear{Aller}{1970}]{b5} Aller M.F., 1970, A\&A, 6, 67
\bibitem[\protect\citeauthoryear{Babel}{1994}]{b6} Babel J., 1994, A\&A, 283, 189
\bibitem[\protect\citeauthoryear{Balona}{1994}]{b7} Balona L.A., 1994, MNRAS, 268, 119
\bibitem[\protect\citeauthoryear{Castelli \& Hubrig}{2004}]{b8} Castelli F., Hubrig S., 2004, A\&A, 425, 263
\bibitem[\protect\citeauthoryear{Cowley et al.}{2006}]{b8} Cowley C.R., Hubrig S., Gonz\'alez G.F., Nu\~nez N., 2006, A\&A, 455, L21
\bibitem[\protect\citeauthoryear{Dolk et al.}{2003}]{b10} Dolk L., Wahlgren G.M., Hubrig S., 2003, A\&A, 402, 299
\bibitem[\protect\citeauthoryear{Dubaj et al.}{2004}]{b11} Dubaj D., Monier R., Alecian G., LeBlanc F., 2004, Semaine de l'Astrophysique Fran\c caise, meeting held in Paris, France, June 14-18, 2004. Edited by F. Combes, D. Barret, T. Contini, F. Meynadier and L. Pagani. Published by EdP-Sciences, Conference Series, 287
\bibitem[\protect\citeauthoryear{Gerbaldi et al.}{1985}]{b12} Gerbaldi M., Floquet M., Hauck M., 1985, A\&A, 146, 341
\bibitem[\protect\citeauthoryear{Grevesse et al.}{2007}]{b13} Grevesse N., Asplund M., Sauval A.J., 2007, SSR, 130, 105
\bibitem[\protect\citeauthoryear{Hauschildt et al.}{1997}]{b14} Hauschildt P.H., Baron E., Allard F., 1997, ApJ, 483, 390
\bibitem[\protect\citeauthoryear{Heacox}{1979}]{b9} Heacox W.D., 1979, AJSS, 41, 675
\bibitem[\protect\citeauthoryear{Holt et al.}{1999}]{b15} Holt R.A., Scholl T.J., Rosner S.D., 1999, MNRAS, 306, 107
\bibitem[\protect\citeauthoryear{Hubrig et al.}{2006}]{b16} Hubrig S., Gonz\'alez, J.F., Savanov, I., et al., 2006, MNRAS, 371, 1953
\bibitem[\protect\citeauthoryear{Hubrig \& Castelli}{2001}]{b17} Hubrig S., Castelli F., 2001, A\&A, 375, 963
\bibitem[\protect\citeauthoryear{Hubrig et al.}{1999}]{b18} Hubrig S., Castelli F., Wahlgren G.M., 1999, A\&A, 346, 139
\bibitem[\protect\citeauthoryear{Hui-Bon-Hoa et al.}{2000}]{b19} Hui-Bon-Hoa A., LeBlanc F., Hauschildt P.H., 2000, ApJ, 535, L43
\bibitem[\protect\citeauthoryear{Jomaron et al.}{1999}]{b20} Jomaron C.M., Dworetsky M.M., Allen C.S., 1999, MNRAS, 303, 555
\bibitem[\protect\citeauthoryear{Khalack et al.}{2008}]{b21} Khalack V.R., LeBlanc F., Behr B.B., Wade G.A., Bohlender D., 2008, A\&A, 477, 641
\bibitem[\protect\citeauthoryear{Khalack et al.}{2007}]{b22} Khalack V.R., LeBlanc F., Bohlender D., Wade G.A., Behr B.B., 2007, A\&A, 466, 667
\bibitem[\protect\citeauthoryear{Khan \& Shulyak}{2007}]{b23} Khan S.A., Shulyak D.V., 2007, A\&A, 469, 1083
\bibitem[\protect\citeauthoryear{Kochukhov et al.}{2005}]{b24} Kochukhov O., Piskunov N., Sachkov M., Kudryavtsev D., 2005, A\&A, 439, 1093
\bibitem[\protect\citeauthoryear{Kupka et al.}{1999}]{b25} Kupka F., Piskunov N., Ryabchikova T.A., Stempels H.C., Weiss W.W., 1999, A\&A, 138, 119
\bibitem[\protect\citeauthoryear{Kurucz}{1994}]{b26} Kurucz R., 1994, CDROM Model Distribution, SAO
\bibitem[\protect\citeauthoryear{Kurucz}{1993}]{b26} Kurucz R., 1993, CDROM Model Distribution, SAO
\bibitem[\protect\citeauthoryear{Landstreet}{1998}]{b27} Landstreet J.D., 1998, A\&A, 338, 1041
\bibitem[\protect\citeauthoryear{Landstreet}{1988}]{b28} Landstreet J.D., 1988, ApJ, 326, 967
\bibitem[\protect\citeauthoryear{Landstreet}{1982}]{b29} Landstreet J.D., 1982, ApJ, 258, 639
\bibitem[\protect\citeauthoryear{Lanz et al.}{1993}]{b30} Lanz T., Artru M.-C., Didelon P., Mathys G., 1993, A\&A, 272, 465
\bibitem[\protect\citeauthoryear{Martin et al.}{1988}]{b31} Martin G.A., Fuhr J.R., Weise W.L., 1988, J. Phys. Chem. Ref. Data, 17, Suppl. 3
\bibitem[\protect\citeauthoryear{Martin et al.}{1985}]{b31} Martin W.C., Zalubas R., Musgrove A., 1985, J. Phys. Chem. Ref. Data, 14, 751
\bibitem[\protect\citeauthoryear{Mathys \& Hubrig}{1995}]{b32} Mathys G., Hubrig S., 1995, A\&A, 293, 810
\bibitem[\protect\citeauthoryear{Michaud}{1970}]{b33} Michaud G., 1970, ApJ, 160, 641
\bibitem[\protect\citeauthoryear{Pickering et al.}{2001}]{1000} Pickering J.C., Thorne A.P., Perez R., 2001, ApJS, 132, 403
\bibitem[\protect\citeauthoryear{Preston}{1974}]{b34} Preston G.W., 1974, ARA\&A, 12, 257
\bibitem[\protect\citeauthoryear{Ralchenko et al.}{2007}]{b35} Ralchenko Yu., Jou F.-C., Kelleher D.E., et al., 2007, NIST Atomic Spectra Database (version 3.1.3)
\bibitem[\protect\citeauthoryear{Ryabchikova et al.}{2005}]{b36} Ryabchikova T.A., Leone F., Kochukhov O., 2005, A\&A, 438, 973
\bibitem[\protect\citeauthoryear{Ryabchikova et al.}{2003}]{b37} Ryabchikova T.A., Wade G.A., LeBlanc F., 2003, Proc. of 210th IAU Symp. on Modelling of Stellar Atmosphres, ed. N. Piskunov, W.W. Weiss, \& D.F. Gray. Published on behalf of the IAU by the Astronomical Society of the Pacific, p.301
\bibitem[\protect\citeauthoryear{Ryabchikova et al.}{1999}]{b38} Ryabchikova T.A., Piskunov N.E., Stempels H.C., Kupka F., Weiss W.W., 1999, Proc. of the 6th International Colloquium on Atomic Spectra and Oscillator Strengths, Victoria BC, Physica Spectra T83, 162
\bibitem[\protect\citeauthoryear{Ryabchikova}{1998}]{b39} Ryabchikova T.A., 1998, Contributions of the Astronomical Observatory Skalnate Pleso, vol. 27, no. 3, 319
\bibitem[\protect\citeauthoryear{Savanov \& Hubrig}{2003}]{b40} Savanov I., Hubrig S., 2003, A\&A, 410, 299
\bibitem[\protect\citeauthoryear{Schaller et al.}{1992}]{b41} Schaller G., Schaerer D., Meynet G., Maeder A., 1992, A\&A, 96, 269
\bibitem[\protect\citeauthoryear{Shorlin et al.}{2002}]{b42} Shorlin S.L.S., Wade G.A., Donati J.-F., et al., 2002, A\&A, 392, 637
\bibitem[\protect\citeauthoryear{Sigut}{2001}]{b43} Sigut T.A.A., 2001, A\&A, 377, L27
\bibitem[\protect\citeauthoryear{Sigut et al.}{2000}]{b44} Sigut T.A.A., Landstreet J.D., Shorlin S.L.S., 2000, ApJ, 530, 89
\bibitem[\protect\citeauthoryear{Smith}{1997}]{b45} Smith K.C., 1997, A\&A, 319, 928
\bibitem[\protect\citeauthoryear{Smith}{1995}]{b46} Smith K.C., 1995, A\&A, 297, 237
\bibitem[\protect\citeauthoryear{Smith \& Dworetsky}{1993}]{b47} Smith K.C., Dworetsky M.M., 1993, A\&A, 274, 335
\bibitem[\protect\citeauthoryear{Smith}{1993}]{b48} Smith K.C., 1993, A\&A, 276, 393
\bibitem[\protect\citeauthoryear{Takada-Hidai}{1991}]{b49} Takada-Hidai M., 1991, Evolution of Stars: The Photospheric Abundance Connection IAUS 145, eds. Michaud G., \& Tutukov A., 137
\bibitem[\protect\citeauthoryear{Wade et al.}{2006}]{b50} Wade G.A., Auri\`ere M., Bagnulo S., Donati J.-F., et al., 2006, A\&A, 451, 293
\bibitem[\protect\citeauthoryear{Wade et al.}{2001}]{b51} Wade G.A., Bagnulo S., Kochukhov O., et al., 2001a, A\&A, 374, 265
\bibitem[\protect\citeauthoryear{Wade et al.}{2001}]{b52} Wade G.A., Ryabchikova T.A., Bagnulo, S., Piskunov, N., 2001b, Magnetic Fields Across the Hertzsprung-Russell Diagram, ed. G. Mathys, S.K. Solanki, \& D.T. Wickramasinghe, ASP Conf. Ser., 248, 341
\bibitem[\protect\citeauthoryear{Wahlgren \& Hubrig}{2000}]{b53} Wahlgren G.M., Hubrig S., 2000, A\&A, 362, L13
\bibitem[\protect\citeauthoryear{White et al.}{1976}]{b54} White R.E., Vaughan A.H., Preston G.W., Swings J.P., 1976, ApJ, 204, 131
\end{thebibliography}
\end{document}